\documentclass[11pt]{article}
\usepackage[letterpaper]{geometry}
\usepackage{amsbsy}
\usepackage{amsmath}
\usepackage{amsthm}
\usepackage{amssymb}
\usepackage{amsfonts}
\usepackage{times}
\usepackage{graphicx}
\usepackage{color}
\usepackage{authblk}
\usepackage{jhtitle}
\usepackage{xspace}
\usepackage{enumitem}

\usepackage[scaled=0.90]{helvet}
\usepackage{ifthen}
\usepackage{xcomment}
\usepackage{versions}
\usepackage{natbib}
\newtheoremstyle{mytheorem}{}{}{\slshape}{}{\bfseries}{}{2mm}{}
\theoremstyle{mytheorem}
\newtheorem{remark}{Remark}
\newtheorem{theorem}{Theorem}

\newtheorem{lemma}{Lemma}

\newcommand{\twoBG}{2BG\xspace}
\newcommand{\threeBG}{3BG\xspace}

\newcommand{\commentt}[1]{}

\newcommand{\bY}{\boldsymbol{Y}}
\newcommand{\bX}{\boldsymbol{X}}
\newcommand{\bbeta}{\boldsymbol{\beta}}
\newcommand{\bi}{\boldsymbol{I}}
\newcommand{\bet}{\boldsymbol{\eta}}

\newcommand{\bA}{\boldsymbol{A}}
\newcommand{\otrans}{\tilde{k}}
\newcommand{\btau}{\boldsymbol{\tau}}
\newcommand{\bgamma}{\boldsymbol{\gamma}}
\newcommand{\bomega}{\boldsymbol{\omega}}
\newcommand{\mc}[2]{\multicolumn{#1}{c}{#2}}

\newcommand{\blue}[1]{{\color{black}{#1}}}

\newcounter{itemnum}

\numberwithin{equation}{section}
\begin{document}

\title{Fast Markov chain Monte Carlo for high dimensional Bayesian regression models
with shrinkage priors}

\author[1]{Rui Jin}
\author[1]{Aixin Tan}
\affil[1]{Department of Statistics and Actuarial Science, University of Iowa}

\date{September 27, 2019}
\date{ }

\maketitle
\renewcommand{\baselinestretch}{1.42}\large\normalsize


\begin{abstract}
 In the past decade, many Bayesian shrinkage models have been developed for linear regression problems where the number of covariates, $p$, is large. 
Computation of the intractable posterior is often done with three-block Gibbs samplers (\threeBG), based on representing the shrinkage priors as scale mixtures of Normal distributions. An alternative computing tool is a state of the art Hamiltonian Monte Carlo (HMC) method, which can be easily implemented in the Stan software. However, we found both existing methods to be inefficient and often impractical for large $p$ problems. Following the general idea of \citet{balaEtal:2018}, we propose two-block Gibbs samplers (\twoBG) for three commonly used shrinkage models, namely, the Bayesian group lasso, the Bayesian sparse group lasso and the Bayesian fused lasso models. We demonstrate with simulated and real data examples that the Markov chains underlying \twoBG's converge much faster than that of \threeBG's, and no worse than that of HMC. At the same time, the computing costs of \twoBG's per iteration are as low as that of \threeBG's, and can be several orders of magnitude lower than that of HMC. As a result, the newly proposed \twoBG is the only practical computing solution to do Bayesian shrinkage analysis for datasets with large $p$. Further, we provide theoretical justifications for the superior performance of \twoBG's. \blue{We establish geometric ergodicity (GE) of Markov chains associated with the \twoBG for each of the three Bayesian shrinkage models. We also prove, for most cases of the Bayesian group lasso and the Bayesian sparse group lasso model, the Markov operators for the \twoBG chains are trace-class. Whereas for all cases of all three Bayesian shrinkage models, the Markov operator for the \threeBG chains are not even Hilbert-Schmidt.}
 \end{abstract}


\renewcommand{\baselinestretch}{1.02}\large\normalsize
\vfil %

\thanks{
\noindent %
}
\renewcommand{\baselinestretch}{1.42}\large\normalsize
\thispagestyle{empty} %
\eject %

\setcounter{page}{1}

\section{Introduction}
\label{sec:intro}
A linear regression model describes how a response variable $y$ is related to potential explanatory variables $x_1,\cdots, x_p$ through a vector of regression coefficients, $\bbeta=(\beta_1, \cdots, \beta_p)$. Sparse estimates of $\bbeta$ are often desired to achieve simple models and stable predictions, especially when the number of explanatory variables $p$ is larger than the sample size $n$. One popular set of tools is the penalized regression method, such that, with proper choice of the penalty function, some $\beta_j$'s will be estimated to be zero. See, e.g., \citet{fan:li:2001}. 
However, it's generally challenging to quantify the uncertainty of the coefficient estimates and that of the resulting sparse models, though there are important recent works in this direction. See, for e.g., \citet{chat:lahi:2011, lock:2014, tibs:2016}.

Alternatively, Bayesian models quantify uncertainties of parameters and latent variables through posterior distributions. Two major types of Bayesian regression models are used to induce sparsity. First, the \textit{spike-and-slab} models assign mixtures of the degenerate distribution at zero and continuous distributions as priors for the $\beta_j$'s, to explore exactly sparse solutions \citep{mitc:beau:88,george:mcculloch:1993}. Such models are natural for sparse problems and enjoy optimality in various senses \citep{john:silv:2004, cast:schm:vand:2015}. But both exact and approximate computing are notoriously difficult for even a moderate number of predictors, as an exploration of $2^p$ sub-models is needed essentially. Models of a second kind, which we will call the Bayesian shrinkage models, assign the $\beta_j$'s continuous prior distributions. Bayesian shrinkage models have gained popularity as their approximate computation are more easily achieved, and many such models lead to good inference results. Some notable examples include the horseshoe priors \citep{carv:pols:scot:2010}, the generalized double Pareto priors
\citep{arma:duns:lee:2013}, and the Dirichlet-Laplace priors \citep{bhat:pati:pill:duns:2015}. 

Bayesian shrinkage methods have been tailored to different regression setups. More often than not, predictors have known structures, such as they form groups, or are ordered in some meaningful way. The Bayesian group lasso and the Bayesian sparse group lasso models are designed for the first situation, while the Bayesian fused lasso model is for the latter. See \citet{xf:ghosh:2015, lee:chen:2015, kyungEtal:2010} for their introduction into the Bayesian framework, and see \citet{ming:yi:2006,huan:breh:ma:2012,noahEtal:2013,robertEtal:2015} for their origins in the penalized regression literature. These Bayesian shrinkage models have been applied to diverse areas such as agriculture, genetics, image processing and ecology. Examples can be found in \cite{hefleyEtal:2017}; \cite{liEtal:2015}; \cite{ramanEtal:2009}; \cite{fan:peng:2016}; \cite{hooten:bobbs:2015}; \cite{greenlawEtal:2017} and \cite{zhangEtal:2014}. Successful applications of the Bayesian group lasso, the Bayesian sparse group lasso and the Baysian fused lasso models depend on fast and reliable computing, which is the pursuit of this paper. 

Since there are usually no closed form expressions for posterior quantities of Bayesian shrinkage models, one would resort to Markov chain Monte Carlo (MCMC) as a most reliable method for approximations. Many shrinkage priors, including the three that we will focus on, enjoy a hierarchical representation. This allows an MCMC scheme called the three-block Gibbs sampler (\threeBG) to explore the posterior. However, \threeBG's often suffer from slow convergence due to high correlation between components of the different blocks, especially that between the regression coefficients in one block and the variance parameters in another \citep{rajar:spar:2015}. An alternative MCMC solution is to use Hamiltonian Monte Carlo (HMC), a general sampling method for exploring continuous distributions. We are not aware of previous usage of HMC to compute the aforementioned Bayesian shrinkage models, but this is easy to set up using the Stan~\citep{carpenterEtal:2017} software, with no need of the hierarchical representation of the prior. Stan implements a certain 
 leapfrog variant of HMC, and is widely-known for producing nearly independent samples in numerous empirical examples, for which standard MCMC algorithms like random walk Metropolis do not mix well. In contrast to the empirical success of HMC, there is limited understanding of its theoretical properties, such as the rate of convergence to the target distribution, and whether the resulting Monte Carlo estimators have finite asymptotic variances \citep{diac:2014}. \blue{Indeed, performance of the HMC heavily depends on 
 the choice of tuning parameters} including the number of leapfrog steps and the step sizes. A rare attempt to identify conditions under which certain variants of HMC converge at a geometric rate is \citet{livi:beta:byrn:giro:2016}, but the conditions therein are non-trivial to check and do not concern exactly the type of HMC implemented in Stan. Despite the lack of theoretical guarantees, we run HMC for the Bayesian shrinkage models in sections~\ref{sec:simulation} and \ref{sec:realdata}. It turns out that, in high-dimensional regression problems with correlated covariates, the computing cost of each iteration of HMC can be very high, while the resulting chains still exhibit moderate to high autocorrelations.

So far, we mentioned that for the three Bayesian shrinkage models of interest, both existing computing solutions, \threeBG and HMC, are far from satisfactory. We develop and study two-block Gibbs samplers (\twoBG) that are reliable and fast in high dimensional cases, where these models are most needed. 
Empirically, we demonstrate that the proposed \twoBG greatly outperforms both \threeBG and HMC in computing efficiency, notably with multiple orders of magnitude improvements in challenging cases where $p$ exceeds $n$. Here, computing efficiency is measured as the effective sample size per unit time, where the effective sample size is inversely proportional to the square root of the asymptotic variance of the Monte Carlo estimator of interest. Looking further into what leads to the improvement, we see that the proposed \twoBG mixes much better than \threeBG in terms of having lag-one autocorrelations closer to 0, with similar computing cost per iteration; and \twoBG mixes similarly to or better than HMC, but costs only a small fraction of the latter per iteration. 

Beyond showing the empirical success of \twoBG's for computing the three Bayesian shrinkage models of interest, we prove in theory that they are reliable, and have superior spectral properties than the corresponding \threeBG's. In more details, there are two commonly used criteria to evaluate Markov chains that explore a target distribution, disregarding computing cost: the rate of convergence (\blue{in total variation distance}) to the target, and the asymptotic variance of Monte Carlo estimators of interest. Concerning convergence rates, it is desirable that a Markov chain has geometric ergodicity (GE), a most regularly used condition to establish the central limit theorem (CLT), which allows standard error evaluations of Monte Carlo estimates. For the proposed \twoBG for each of the three variants of the Bayesian lasso model, 
we establish GE of the respective Markov chain. 
In addition, both the convergence rate of a Markov chain and asymptotic variances of Monte Carlo estimators depend on the spectrum of the corresponding Markov operator. See, e.g., \citet{rose:2015} and \citet{mira:2001}. \blue{For the Bayesian group lasso and the Bayesian sparse group lasso models, we are able to prove for almost all cases that the Markov operators associated with \twoBG are trace-class whereas that of \threeBG are not even Hilbert-Schmidt. According to the definition of these properties that we provide in section~2 of the supplement, this means that the (self-adjoint) \twoBG Markov operator has 
eigenvalues that are absolutely summable, and hence square-summable (as the eigenvalues are bounded above by $1$ in absolute value). While for the \threeBG chain, the absolute value of its operator has non-square-summable eigenvalues. 
 In other words, the \twoBG Markov operators have qualitative smaller spectrum than that of \threeBG, and the former are expected to enjoy faster convergence rate, which agree with our empirical observations.}


Another benefit of showing the trace class results for \twoBG is to provide theoretical foundations for further acceleration of these algorithms. Indeed, \twoBG is a kind of Data Augmentation (DA) algorithm. There is a well-studied ``sandwich" trick \citep{liu:wu:1999, meng:vand:1999, hobert:Marchev:2008} that allows many ways to insert an extra step in each iteration of a DA, to achieve potentially much faster convergence rate. The sandwich trick will improve the computational efficiency of DA provided the extra step is inexpensive, and the trick has seen many successful implementations \citep{liu:wu:1999, meng:vand:1999, vand:meng:2001, marc:hobe:2004, hobe:roy:robe:2011, ghos:tan:2015}. The design of good sandwich algorithms for the Bayesian shrinkage models is beyond the scope of this paper. Nevertheless, our results that certain \twoBG samplers are trace class imply that any sandwich improvements of them are also trace-class and have better spectral properties, in the sense that eigenvalues of the improved algorithms are dominated by that of the basic \twoBG \citep{khar:hobe:2011}. 

The \twoBG algorithms proposed and analyzed in this paper stem from the work of \citet{balaEtal:2018}, which introduced the two-block strategy for a class of Bayesian models for regression. In the case of the Bayesian lasso model, they analyzed the convergence rates and the spectral properties for the corresponding \twoBG and \threeBG, and acknowledged that such a theoretical study is ``a big and challenging undertaking". A main contribution of our work is to develop these theoretical results for the aforementioned three variants of the Bayesian lasso models, which complement and support the empirical advantage we observed of using \twoBG over other existing computing solutions. 

\blue{The rest of the paper is organized as follows. In section~\ref{sec:bsm}, we present the three Bayesian shrinkage models and review the existing \threeBG algorithms that explore their posterior distributions. In section~\ref{sec:fast}, we propose the \twoBG algorithms for these models, establish geometric ergodicity of the underlying Markov chains and show they have superior spectral properties than their \threeBG counterparts.  In sections~\ref{sec:simulation} and \ref{sec:realdata}, we perform empirical comparisons among HMC, \threeBG and \twoBG in simulation studies and real data applications.}

\section{Bayesian Shrinkage Models and Standard Three-block Gibbs Samplers}
\label{sec:bsm}
Let $\bY \in \mathbb{R}^n$ be the response vector, $\bX$ be the $n \times p$ design matrix, $\bbeta \in \mathbb{R}^p$ be the vector of  regression coefficients and $\sigma^2 > 0 $ be the residual variance in a regression problem. We consider a general framework of Bayesian models as follows: 
\begin{equation}\label{bsm}
\begin{split}
\bY | \bbeta, \sigma^2 &  \sim  	{\cal N}_n \left(\bX \bbeta,\sigma^2 \bi_n \right),  \\
\bbeta | \bet, \sigma^2  & \sim   {\cal N}_p \left(\mathbf{0},\sigma^2 \Sigma_{\bet} \right), \\
\bet & \sim  p(\bet),  \\
\sigma^2 & \sim  \text{Inverse-Gamma}(\alpha, \xi)\,.
\end{split}
\end{equation}
Here, $\Sigma_{\bet}$ is a $p \times p$ positive definite matrix indexed by $\bet$, with a proper prior $p(\bet)$ on $\bet$. \blue{The hyperparameters $\alpha$ and $\xi$ are non-negative real numbers specified by users.}
Many Bayesian shrinkage models can be represented using \eqref{bsm}, e.g., those under the lasso prior \citep{park:case:2008}, the enet prior \citep{li:lin:2010}, the horseshoe prior \citep{carv:pols:scot:2010}, and the Dirichlet--Laplace prior \citep{bhat:pati:pill:duns:2015}. Bayesian inference of the model is based on the joint posterior distribution of $(\bbeta, \sigma^2, \bet)$, especially the $(\bbeta, \sigma^2)-$marginal and various summaries of it. These distributions are analytically intractable and difficult to draw independent samples from.  \blue{The focus of this paper is to find efficient MCMC algorithms to sample from the joint posterior distribution of $(\bbeta, \sigma^2)$ for the model in (\ref{bsm}).} 
An alternative solution through variational approximation is usually of much lower accuracy (see, e.g., \citet{nevi:orme:wand:2014}), and not discussed here.  

Note that the model in (\ref{bsm}) yields the following conditional distributions:
\begin{equation}\label{three_cond}
\begin{split}
	\bet | \bbeta, \sigma^2, \bY  & \sim  \pi \left(	\bet | \bbeta, \sigma^2, \bY \right),  \\
\sigma^2 | \bbeta, \bet, \bY &\sim   \text{Inverse-Gamma} \left(\frac{n+p+2\alpha}{2}, \frac{\|\bY - \bX \bbeta \|_2^2 + \bbeta^T \Sigma_{\bet}^{-1} \bbeta + 2\xi}{2} \right), \\
\bbeta | \sigma^2, \bet, \bY & \sim  {\cal N}_p \left(\bA^{-1}_{\bet}\bX^T\bY, \sigma^2 \bA^{-1}_{\bet} \right),
\end{split}
\end{equation}
where $\bA_{\bet} = \bX^T\bX + \Sigma_{\bet}^{-1}$. Hence, one can use a Gibbs sampler to update the three blocks of components, $\bet$, $\sigma^2$ and $\bbeta$ in turn, as long as $ \pi\left(\bet | \bbeta, \sigma^2, \bY\right)$ is easy to sample from. \blue{Such Gibbs samplers have been adopted in \cite{kyungEtal:2010}; \cite{grifEtal:2010}; \cite{palEtal:2017} and \cite{li:lin:2010} to study special cases of model~\eqref{bsm}.  Note that $(\bbeta, \sigma^2)$ are the parameters of main interest in the Bayesian regression models, and the $(\bbeta, \sigma^2)$-marginal of the Gibbs sampler, $\tilde{\boldsymbol{\Phi}}:= \left\{(\tilde{\bbeta}_m, \tilde{\sigma}_m^2) \right\}_{m=0}^{\infty}$ is itself a Markov chain on the state space $\mathbb{R}^p \times \mathbb{R}_{+}$. This Markov chain is what we will refer to as the \threeBG in this paper, with Markov transition density:}
\begin{eqnarray}\label{old_tran}
\blue{
\otrans\left[(\tilde{\bbeta}_0, \tilde{\sigma}_0^2), (\tilde{\bbeta}_1, \tilde{\sigma}_1^2) \right] = \int_{\mathbb{R}^s_{+}} \pi(\tilde{\bbeta}_1 | \tilde{\bet}, \tilde{\sigma}_1^2, \bY) \pi(\tilde{\sigma}_1^2 | \tilde{\bbeta}_0, \tilde{\bet}, \bY) \pi(\tilde{\bet} | \tilde{\bbeta}_0, \tilde{\sigma}_0^2, \bY) d\tilde{\bet}.}
\end{eqnarray} 
We next review three special cases of model~\eqref{bsm} that are useful for different regression setups. 
 
\subsection{The Bayesian Group Lasso}
\label{sec:bgl}
In regression problems, explanatory variables often occur in groups. Grouping structures can be implied by knowledge of the area of application. For instance, in modeling health of individuals, variables may belong to different categories such as social and economic environment, physical environment, and individual characteristics. Also, quantitative variables may enter a model as groups of basis functions, and qualitative variables with multiple levels could be represented by groups of dummy variables. See \citet{huan:breh:ma:2012} for an extensive review of problems that involve groups of variables. Suppose there are $K$ groups of explanatory variables. For $k=1,\ldots, K$, let $m_k$ denote the size of the $k$th group, $\bbeta_{G_k}$ the $m_k$-dim vector of regression coefficients, and $\bX_{G_k}$ the corresponding sub-design matrices. \cite{ming:yi:2006} presented the group lasso method for estimating $\bbeta$ as follows, 
\begin{eqnarray}\label{gl}
	\hat{\bbeta}_{\text{group}} = \arg\min_{\bbeta} \left\|\bY - \sum_{k=1}^K \bX_{G_k} \bbeta_{G_k} \right\|_2^2 + \lambda \sum_{k=1}^K  \left\|\bbeta_{G_k} \right\|_2\,,
\end{eqnarray}
\blue{where the penalty parameter $\lambda$} decides the intensity of the shrinkage and is often tuned by cross-validation. An obvious Bayesian analogue of the above method can be obtained by combining the first line of \eqref{bsm} and the prior
 \begin{eqnarray}\label{prior_bgl}
\pi(\bbeta | \sigma^2) \propto \exp \left(-\frac{\lambda}{\sigma} \sum_{k = 1}^K \| \bbeta_{G_k} \|_2 \right)\,.
\end{eqnarray}
  \cite{kyungEtal:2010} showed a hierarchical representation of \eqref{prior_bgl}, and the model is completed with a prior on $\sigma^2$:
\begin{equation}\label{bgl_frame}
 \begin{split}
	\bY | \bbeta, \sigma^2 & \sim  {\cal N}_n (\bX \bbeta, \sigma^2 \bi_n),  \\
\bbeta_{G_k} | \tau_k^2, \sigma^2 & \overset{ind}{\sim} {\cal N}(\mathbf{0}_{m_k}, \sigma^2\tau_k^2 \bi_{m_k}), \hspace{.2cm} k = 1,\dots, K,  \\
\tau_k^2 & \overset{ind}{\sim} \text{Gamma}\left(\frac{m_k + 1}{2}, \frac{\lambda^2}{2} \right), \hspace{.2cm} k = 1,\dots, K,  \\
\sigma^2 & \sim  \text{Inverse-Gamma}(\alpha, \xi)\,.
 \end{split}
 \end{equation}
Here, $\alpha, \xi \geq 0$ and $\lambda > 0$ are user specified hyperparameters. Note that \eqref{bgl_frame} is indeed a special case of \eqref{bsm}, with $\boldsymbol{\eta}$ \blue{specified to be} $\btau = (\tau_1, \dots, \tau_K)$, and $\boldsymbol{\Sigma_\eta}$ \blue{specified to be} 
\begin{eqnarray*}
\boldsymbol{D_\tau} = \text{diag}(\underbrace{\tau_1^2, \dots, \tau_1^2}_{m_1}, \underbrace{\tau_2^2, \dots, \tau_2^2}_{m_2}, \dots, \underbrace{\tau_K^2, \dots, \tau_K^2}_{m_K}).
\end{eqnarray*}

\blue{For the model in (\ref{bgl_frame}), derivations of the full conditionals of $\btau^2$, $\sigma^2$  and $\bbeta$ are in the supplementary material.
These conditionals enable a Gibbs sampler to explore the corresponding posterior, which updates the three blocks $\btau^2$, $\sigma^2$ and $\bbeta$ in that order. 
The $(\sigma^2, \bbeta)$-marginal of this Gibbs sampler is a Markov chain that we call the \threeBG chain for the Bayesian group lasso model. Its transition density is given by
\begin{eqnarray}\label{old_tran_gl}
\otrans_{gl} \left[(\tilde{\bbeta}_0, \tilde{\sigma}_0^2), (\tilde{\bbeta}_1, \tilde{\sigma}_1^2) \right]= \int_{\mathbb{R}^K_{+} }  \pi(\tilde{\bbeta}_1 | \tilde{\sigma}_1^2,\tilde{\btau}^2,  \bY) \pi(\tilde{\sigma}_1^2 | \tilde{\bbeta}_0, \tilde{\btau}^2, \bY) \pi (\tilde{\btau}^2 | \tilde{\bbeta}_0, \tilde{\sigma}_0^2, \bY) d\tilde{\btau}^2.
\end{eqnarray}}

\subsection{The Bayesian Sparse Group Lasso}
\label{sec:bsgl}
In the same setup of the previous section where the explanatory variables form $K$ groups, \cite{noahEtal:2013} introduced the sparse group lasso method that promotes sparsity in regression coefficients both among groups and within each group. For penalty parameters $\lambda_1, \lambda_2 > 0$, the sparse group lasso estimator is defined as
\begin{eqnarray}\label{sgl}
	\hat{\bbeta}_{\text{sgroup}} = \arg\min_{\bbeta} \left\|\bY - \sum_{k=1}^K \bX_{G_k} \bbeta_{G_k} \right\|_2^2  + \lambda_1 \sum_{k=1}^K \left\|\bbeta_{G_k} \right\|_2 + \lambda_2 \left\|\bbeta \right\|_1\,,
\end{eqnarray} 
where $\lambda_1$ and $\lambda_2 $ are often tuned by cross validation. 
A Bayesian analogue of \eqref{sgl} can be obtained by specifying  
  \begin{eqnarray}\label{prior_bsgl}
\pi(\bbeta | \sigma^2) \propto \exp \left(-\frac{\lambda_1}{\sigma} \sum_{k = 1}^K \| \bbeta_{G_k} \|_2  - \frac{\lambda_2}{\sigma}\|\bbeta \|_1 \right)\,.
\end{eqnarray}
We will refer to the model that uses \eqref{prior_bsgl} and completed with an Inverse-Gamma prior on $\sigma^2$ as the Bayesian sparse group lasso model \citep{xf:ghosh:2015}. \blue{For $k = 1,\dots, K$,  let
\begin{eqnarray*}
	\boldsymbol{V}_k = \text{diag} \left\{\left(\frac{1}{\tau_k^2} + \frac{1}{\gamma_{k, j}^2}\right)^{-1} \vspace{.2cm} : j = 1,\dots, m_k \right\}\,,
\end{eqnarray*}
and 
\begin{eqnarray*}
	\pi_k \propto \prod_{j = 1}^{m_k} \left[(\gamma_{k, j}^2)^{-\frac{1}{2}} \left(\frac{1}{\gamma_{k, j}^2} + \frac{1}{\tau_k^2} \right)^{-\frac{1}{2}} \right] (\tau_k^2)^{-\frac{1}{2}} \exp \left(-\frac{\lambda_2^2}{2} \sum_{j=1}^{m_k}\gamma_{k, j}^2 - \frac{\lambda_1^2}{2} \tau_k^2 \right)\,.
\end{eqnarray*}}
\blue{The Bayesian sparse group lasso model} is a special case of the Bayesian shrinkage model \eqref{bsm}, as it has the following hierarchical representation:
\begin{equation}\label{bsgl_frame}
\begin{split}
\bY | \bbeta, \sigma^2 & \sim  {\cal N}_n (\bX \bbeta, \sigma^2 \bi_n),  \\
\bbeta_{G_k} | \btau^2, \bgamma^2, \sigma^2 & \overset{ind}{\sim} {\cal N}(\mathbf{0}_{m_k}, \sigma^2\boldsymbol{V}_k), \hspace{.2cm} k = 1,\dots, K,  \\
(\gamma_{k, 1}^2, \dots, \gamma_{k, m_k}^2, \tau_k^2)& \overset{ind}{\sim} \pi_k, \hspace{.2cm} k = 1,\dots, K,  \\
\sigma^2 & \sim  \text{Inverse-Gamma}(\alpha, \xi),
\end{split}
\end{equation}
where $\bgamma^2 = (\gamma_{1, 1}^2, \dots, \gamma_{1, m_1}^2, \dots, \gamma_{K, 1}^2, \dots, \gamma_{K, m_K}^2)$, $\btau^2 = (\tau_1^2, \dots, \tau_K^2)$, $\alpha, \xi \geq 0$, and $\lambda_1, \lambda_2 > 0$ are user specified hyperparameters.

\blue{For the model in (\ref{bsgl_frame}), the full conditionals of $(\btau^2, \bgamma^2)$, $\bbeta$ and $\sigma^2$ are available in the supplementary material. 
Hence a Gibbs sampler can be run that updates the three blocks $(\btau^2, \bgamma^2)$, $\sigma^2$, and $\bbeta$, in that order. 
The $(\sigma^2, \bbeta)$-marginal forms a Markov chain that we call the \threeBG chain for the Bayesian sparse group lasso model. Its transition density is given by 
\begin{eqnarray}\label{old_tran_sgl}
\otrans_{sgl} \left[(\tilde{\bbeta}_0, \tilde{\sigma}_0^2), (\tilde{\bbeta}_1, \tilde{\sigma}_1^2) \right]= \int_{\mathbb{R}^K_{+} } \int_{\mathbb{R}^p_{+} }  \pi(\tilde{\bbeta}_1 |\tilde{\sigma}_1^2,  \tilde{\btau}^2, \tilde{\bgamma}^2,\bY) \pi(\tilde{\sigma}_1^2 | \tilde{\bbeta}_0, \tilde{\btau}^2, \tilde{\bgamma}^2,\bY) \pi (\tilde{\btau}^2, \tilde{\bgamma}^2 | \tilde{\bbeta}_0, \tilde{\sigma}_0^2, \bY) d\tilde{\bgamma}^2d\tilde{\btau}^2.  \hspace{.5cm}
\end{eqnarray}}

\subsection{The Bayesian Fused Lasso}
\label{sec:bfl}
For regression problems in which explanatory variables can be ordered in some meaningful way, \cite{robertEtal:2015} developed the fused lasso method, that penalizes the $\ell_1$ norm of both the coefficients and their successive differences. For fixed tuning parameters $\lambda_1, \lambda_2 >0 $, the fused lasso estimator is defined as
\begin{eqnarray}\label{fl}
	\hat{\bbeta}_{\text{fused}} = \arg\min_{\bbeta} \|\bY - \bX \bbeta \|_2^2 + \lambda_1 \|\bbeta \|_1 + \lambda_2 \sum_{j = 1}^{p-1} |\beta_{j+1} - \beta_j|.
\end{eqnarray}
We now present the Bayesian counterpart of the fused lasso method. Let the conditional prior of $\bbeta | \sigma^2$ be 
\begin{eqnarray}\label{prior_bfl}
\pi(\bbeta | \sigma^2) \propto \exp \left(- \frac{\lambda_1}{\sigma}\|\bbeta \|_1  - \frac{\lambda_2}{\sigma} \sum_{j = 1}^{p-1} |\beta_{j+1} - \beta_j| \right).
\end{eqnarray}
\blue{
Let $\boldsymbol{\Sigma}_{\tau, \omega}$ be a matrix  such that 
\begin{eqnarray*}
	\boldsymbol{\Sigma}_{\tau, \omega}^{-1} = \begin{bmatrix}
		\frac{1}{\tau_1^2} + \frac{1}{\omega_1^2} & - \frac{1}{\omega_1^2} & \cdots & 0\\
		- \frac{1}{\omega_1^2} & \frac{1}{\tau_2^2} + \frac{1}{\omega_2^2}  + \frac{1}{\omega_1^2} & \cdots & 0\\
		0 & -\frac{1}{\omega_2^2}  & \cdots & 0\\
		\cdots & \cdots & \ddots & \cdots \\
		0 & 0 &  \frac{1}{\tau_{p-1}^2} + \frac{1}{\omega_{p-2}^2}  + \frac{1}{\omega_{p-1}^2}  & -\frac{1}{\omega_{p-1}^2}\\
		0 & 0 & -\frac{1}{\omega_{p-1}^2} &  \frac{1}{\tau_{p}^2} + \frac{1}{\omega_{p-1}^2} 
	\end{bmatrix}.
\end{eqnarray*}}

\blue{A full Bayesian fused lasso model based on a hierarchical representation of (\ref{prior_bfl}) and a prior on $\sigma^2$ was presented in \cite{vats:2017}, as shown below}:
\begin{equation}\label{bfl_frame}
\begin{split}
\bY | \bbeta, \sigma^2 & \sim  {\cal N}_n(\bX\bbeta, \sigma^2 \bi_n),  \\
\bbeta | \btau^2, \bomega^2, \sigma^2 & \sim  {\cal N}_p(\mathbf{0}_p, \sigma^2 \boldsymbol{\Sigma}_{\tau, \omega}),  \\
\blue{\pi(\btau^2, \bomega^2)} & \blue{\propto} \blue{\text{ det}(\boldsymbol{\Sigma}_{\tau, \omega})^{1/2}} \blue{\left(\prod_{i = 1}^p (\tau_i^2)^{-1/2} e^{-\lambda_1\tau_i^2/2} \right) \left(\prod_{i = 1}^{p - 1} (\omega_i^2)^{-1/2} e^{-\lambda_2\omega_i^2/2} \right) } \\
\sigma^2 & \sim  \text{Inverse- Gamma}(\alpha, \xi),
\end{split}
\end{equation}
where $\btau = (\tau_1, \dots, \tau_p)$ and $\bomega = (\omega_1, \dots, \omega_{p-1})$ play the role of $\eta$ of \eqref{bsm}, and $\alpha, \xi \geq 0$, and $\lambda_1, \lambda_2 > 0$ are hyperparameters specified by the user. 
\blue{The full conditionals  of $(\btau^2, \bomega^2)$, $\bbeta$ and $\sigma^2$ can be derived from   (\ref{bfl_frame}) and are included in the supplementary material.
These conditionals enable a Gibbs sampler that updates the three blocks $(\btau^2, \bomega)$, $\sigma^2$  and $\bbeta$ in that order. The $(\sigma^2, \bbeta)$-marginal forms a Markov chain with transition density 
\begin{eqnarray}\label{old_tran_fl}
\otrans_{fl} \left[(\tilde{\bbeta}_0, \tilde{\sigma}_0^2), (\tilde{\bbeta}_1, \tilde{\sigma}_1^2) \right]= \int_{\mathbb{R}^p_{+} }  \int_{\mathbb{R}^{p-1}_{+} }  \pi(\tilde{\bbeta}_1 | \tilde{\sigma}_1^2, \tilde{\btau}^2, \tilde{\bomega}^2, \bY) \pi(\tilde{\sigma}_1^2 | \tilde{\bbeta}_0, \tilde{\btau}^2, \tilde{\bomega}^2,\bY) \pi (\tilde{\btau}^2, \tilde{\bomega}^2 | \tilde{\bbeta}_0, \tilde{\sigma}_0^2, \bY) d\tilde{\bomega}^2d \tilde{\btau}^2\,.  \hspace{.5cm}
\end{eqnarray}
We refer to the above as the \threeBG chain for the Bayesian fused lasso model.}

\section{Fast MCMC for Bayesian Shrinkage Models}
\label{sec:fast}
The \threeBG algorithms in section~\ref{sec:bsm} are the most commonly used computing solution for the aforementioned Bayesian shrinkage models. Despite straightforward implementations, \threeBG converges slowly in high-dimensional settings. For a special case of the Bayesian shrinkage model in (\ref{bsm}) with $\alpha= \xi =0$, which assigns a ``non-informative"  improper prior on $\sigma^2$, 
\cite{balaEtal:2018} noted that the conditional distribution $\pi(\sigma^2 | \bet, \bY)$ is also tractable, and inexpensive to sample from. Hence, $(\bbeta, \sigma^2)$ can be updated in one block by drawing from $\pi(\sigma^2 | \bet, \bY)$ and then $\pi(\bbeta | \sigma^2, \bet, \bY)$. 
We now slightly generalize Lemma~1 of~\cite{balaEtal:2018} by developing a tractable form of the conditional posterior distribution $\pi(\sigma^2 | \bet, \bY)$ for the general Bayesian shrinkage model in (\ref{bsm}), where the prior can be either improper or proper. The proof can be found in the supplementary materials. 
\begin{lemma}\label{cond_sigma}
 For the Bayesian shrinkage model in (\ref{bsm}),  suppose that the prior on $\sigma^2$ is $ \text{Inverse-Gamma}(\alpha, \xi)$ for $\alpha,\xi \geq 0$. Then, $\sigma^2 | \bet, \bY$ follows an Inverse-Gamma distribution with shape parameter $\frac{n}{2} + \alpha$ and scale parameter $\bY^T(\bi - \bX(\bX^T\bX + \boldsymbol{\Sigma}_{\bet}^{-1} )^{-1}\bX^T)\bY / 2 + \xi$.
\end{lemma}

\blue{Lemma~\ref{cond_sigma} leads to a Gibbs sampler that updates the two blocks $\bet$ and $(\sigma^2, \bbeta)$ in turn:}
\begin{equation}\label{2gs_general}
\begin{split}
\bet | \bbeta, \sigma^2, \bY &\sim \pi(\bet | \bbeta, \sigma^2, \bY), \\
(\sigma^2, \bbeta) | {\bet}, \bY& \begin{cases} 
\sigma^2 | \bet, \bY \sim \text{Inverse-Gamma} \left(\frac{n}{2} + \alpha, \frac{\bY^T(\mathbf{I}_n - \bX(\bX^T\bX + \boldsymbol{\blue{\Sigma}}_{\eta}^{-1})^{-1}\bX^T)\bY}{2} + \xi \right),\\
& \\
\bbeta | \sigma^2, \bet, \bY \sim {\cal N}_p \left(\left(\bX^T\bX + \boldsymbol{\blue{\Sigma}}_{\eta}^{-1}\right)^{-1}\bX^T\bY, \sigma^2 \left(\bX^T\bX + \boldsymbol{\blue{\Sigma}}_{\eta}^{-1}\right)^{-1} \right).
\end{cases}
\end{split}
\end{equation}
\blue{Comparing \eqref{2gs_general} to \eqref{three_cond}, note that the only technical difference between \twoBG and its \threeBG counterpart lies in the step of drawing $\sigma^2$, which has similar cost for the two algorithms. Note that despite a seemingly harder-to-evaluate conditional distribution of $\sigma^2$ in \twoBG that requires the inversion of a $p \times p$ matrix $\bA_{\bet} = \bX^T\bX + \Sigma_{\bet}^{-1}$, this quantity is actually needed in other steps of both algorithms, hence not an extra burden for \twoBG.}

\blue{Convergence properties of the Markov chain that underlies this Gibbs sampler is characterized by its $\left(\bbeta,  \sigma^2\right)$-marginal, $\boldsymbol{\Phi}:= \left\{({\bbeta}_m, {\sigma}_m^2) \right\}_{m=0}^{\infty}$. We call this Markov chain on the state space $\mathbb{R}^p \times \mathbb{R}_{+}$ the \twoBG chain. Its transition density is given by}
\begin{eqnarray}\label{new_tran}
	k\left[(\bbeta_0, \sigma_0^2), (\bbeta_1, \sigma_1^2) \right] = \int_{\mathbb{R}^s_{+}} \pi(\bbeta_1 |  \sigma_1^2, \bet, \bY) \pi(\sigma_1^2 |  \bet, \bY) \pi(\bet | \bbeta_0, \sigma_0^2, \bY) d\bet.
\end{eqnarray}

\blue{ In the subsections below, we provide details of the new \twoBG samplers for the three aforementioned Bayesian shrinkage models, each followed by a theorem on their theoretical properties and that of their \threeBG counterparts. Proofs of the three Theorems are in the supplementary materials, and we give a brief overview of them here. For Bayesian group lasso models and Bayesian sparse group lasso models with hyperparameters $\alpha \geq 0$ and $\xi > 0$, we show that \twoBG chains are trace-class, which imply they are Hilbert-Schmidt and have GE. For these two Bayesian models with hyperparameters $\alpha \geq 0$ and $\xi =0$, and for the Bayesian fused lasso models with $\alpha, \xi \geq 0$ we do not yet know if their \twoBG chains are trace-class, due to technical difficulties; instead, we establish their GE directly, using the classical drift and minorization condition \citep{rose:1995}. }
\blue{\begin{remark}
	It's straightforward from their definitions that a self-adjoint trace-class operator is also Hilbert-Schmidt. Further, a Hilbert-Schmidt operator is always compact. For a Harris ergodic Markov chain, the compactness implies that the spectral radius of the Markov operator is strictly less than $1$,  hence geometrically ergodic. See, e.g., \cite{chan:geye:1994} and \cite{robe:rose:1997}. Note that the Markov transition densities~\eqref{trans_2bgl}, \eqref{trans_2bsgl} and~\eqref{trans_2bfl} are indeed strictly positive, it follows that the corresponding \twoBG chains are Harris ergodic \citep[Lemma~1]{tan:hobe:2009}. 
Therefore, the Theorems proved in this paper show that all \twoBG chains for the three types of Bayesian shrinkage models have GE, while many of them are also shown to have the stronger trace-class property. In comparison, \threeBG chains for the same Bayesian shrinkage models were shown to have GE by \citet{vats:2017}, but our Theorems show that none of them are Hilbert-Schmidt. 
\end{remark}
}

\subsection{The Two-block Gibbs Sampler for the Bayesian Group Lasso}
\label{sec:2gs_bgl}
For the Bayesian group lasso model, the \twoBG updates described in (\ref{2gs_general}) become \begin{equation}\label{2gs_bgl}
\begin{split}
	\frac{1}{\tau_k^2} | \bbeta, \sigma^2, \bY &\overset{ind}{\sim}  \text{Inverse-Gaussian}\left(\sqrt{\frac{\lambda^2 \sigma^2}{\|\bbeta_{G_K} \|_2^2}} , \lambda^2 \right), \text{ for } k = 1, \dots, K,  \\
(\sigma^2, \bbeta) | {\btau^2}, \bY& \begin{cases} 
\sigma^2 | \btau^2, \bY  \sim \text{Inverse-Gamma} \left(\frac{n}{2} + \alpha, \frac{\bY^T(\mathbf{I}_n - \bX(\bX^T\bX + \boldsymbol{D}_{\tau}^{-1})^{-1}\bX^T)\bY}{2} + \xi \right),\\
 \\
\bbeta | \sigma^2, \btau^2, \bY  \sim {\cal N}_p \left(\left(\bX^T\bX + \boldsymbol{D}_{\tau}^{-1}\right)^{-1}\bX^T\bY, \sigma^2 \left(\bX^T\bX + \boldsymbol{D}_{\tau}^{-1}\right)^{-1} \right)\,.
\end{cases}
\end{split}
\end{equation}
The one-step transition density of the $\left(\bbeta,  \sigma^2\right)$-chain is given by
\begin{eqnarray}\label{trans_2bgl}
	k_{gl} \left[(\bbeta_0, \sigma^2_0), (\bbeta_1, \sigma^2_1) \right] = \int_{\mathbb{R}^K_{+}} \pi(\bbeta_1 | \sigma_1^2, \btau^2, \bY) \pi(\sigma_1^2 | \btau^2, \bY)   \pi (\btau^2 | \bbeta_0, \sigma_0^2, \bY) d\btau^2. 
\end{eqnarray}

\begin{theorem}\label{bgl_trace}
\blue{For the Bayesian group lasso model with hyperparameters $\alpha \geq 0$ and $\xi >0$, the Markov operator of the \twoBG chain with transition density $k_{gl}$ is trace-class, and hence is Hilbert-Schmidt; while that of  the \threeBG chain with transition density $\tilde{k}_{gl}$ is not Hilbert-Schmidt for any $\alpha, \xi \geq 0$. In the case of $\alpha \geq 0$ and $\xi = 0$, the \twoBG chain is geometrically ergodic.}
\end{theorem}

\subsection{The Two-block Gibbs Sampler for the Bayesian Sparse Group Lasso}
\label{sec:2gs_bsgl}
For the Bayesian sparse group lasso model, the \twoBG updates the two blocks $(\btau, \bgamma)$ and $(\sigma^2, \bbeta)$ in each iteration as follows:
\begin{equation}\label{2gs_bsgl}
\begin{split}
\frac{1}{\tau_k^2} | \bbeta, \sigma^2, \bY &\overset{ind}{\sim}  \text{Inverse-Gaussian}\left(\sqrt{\frac{\lambda_1^2 \sigma^2}{\|\bbeta_{G_K} \|_2^2}} , \lambda_1^2 \right), \text{ for $k = 1, \dots, K$, and independently, }\\
\frac{1}{\gamma_{k, j}^2}  | \bbeta, \sigma^2, \bY &\overset{ind}{\sim} \text{Inverse-Gaussian} \left(\sqrt{\frac{\lambda_2^2 \sigma^2}{\beta_{k, j}^2}}, \lambda_2^2 \right), \hspace{.1cm}  \text{ for } k =1, \dots, K  \text{ and } j = 1,\dots, m_k, \\
(\sigma^2, \bbeta) | {\btau^2, \bgamma^2, \bY}&\sim \begin{cases} 
\sigma^2 | \btau^2, \bgamma^2,\bY  \sim \text{Inverse-Gamma} \left(\frac{n}{2} + \alpha, \frac{\bY^T(\mathbf{I}_n - \bX(\bX^T\bX + \boldsymbol{V}_{\tau, \gamma}^{-1})^{-1}\bX^T)\bY}{2} + \xi \right),\\
& \\
\bbeta | \sigma^2, \btau^2, \bgamma^2,\bY \sim {\cal N}_p \left(\left(\bX^T\bX + \boldsymbol{V}_{\tau, \gamma}^{-1}\right)^{-1}\bX^T\bY, \sigma^2 \left(\bX^T\bX + \boldsymbol{V}_{\tau, \gamma}^{-1}\right)^{-1} \right)\,.
\end{cases}
\end{split}
\end{equation}
The one-step transition density  of the $\left(\bbeta,  \sigma^2\right)$-chain is given by
\begin{eqnarray}\label{trans_2bsgl}
k_{sgl} \left[(\bbeta_0, \sigma^2_0), (\bbeta_1, \sigma^2_1) \right] = \int_{\mathbb{R}^K_{+}} \int_{\mathbb{R}^p_{+}}  \pi(\bbeta_1 | \sigma_1^2, \btau^2, \bgamma^2, \bY)  \pi(\sigma_1^2 | \btau^2, \bgamma^2, \bY)  \pi (\btau^2, \bgamma^2 | \bbeta_0, \sigma_0^2, \bY) d\btau^2 d\bgamma^2. \hspace{.5cm}
\end{eqnarray}

\begin{theorem}\label{bsgl_trace}
\blue{For the Bayesian sparse group lasso model with hyperparameters $\alpha \geq 0$ and $\xi >0$, 
the Markov operator of the  \twoBG chain with transition density $k_{sgl}$ is trace-class, and hence is Hilbert-Schmidt; while that of the \threeBG chain with transition density $\tilde{k}_{sgl}$ is not Hilbert-Schmidt for any $\alpha, \xi \geq 0$. In the case of $\alpha \geq 0$ and $\xi = 0$, the \twoBG chain is geometrically ergodic.}
\end{theorem}

\subsection{The Two-block Gibbs Sampler for the Bayesian Fused Lasso}
\label{sec:2gs_bfl}
For the Bayesian fused lasso model, the \twoBG updates the two blocks $(\btau, \bomega)$ and $(\sigma^2, \bbeta)$ in each iteration as follows:
\begin{equation}\label{2gs_bfl}
\begin{split}
\frac{1}{\tau_j^2} | \bbeta, \sigma^2, \bY &\overset{ind}{\sim}  \text{Inverse-Gaussian}\left(\sqrt{\frac{\lambda_1^2 \sigma^2}{\beta_j^2}} , \lambda_1^2 \right), \text{ for  $j = 1, \dots, p,$ and independently,} \\
\frac{1}{\omega_{ j}^2}  | \bbeta, \sigma^2, \bY &\overset{ind}{\sim} \text{Inverse-Gaussian} \left(\sqrt{\frac{\lambda_2^2 \sigma^2}{(\beta_{j + 1} - \beta_j)^2}}, \lambda_2^2 \right), \hspace{.1cm} \text{ for } j =1, \dots, p-1 \\
(\sigma^2, \bbeta) | {\btau^2, \bomega^2, \bY}& \begin{cases} 
\sigma^2 | \btau^2, \bomega^2,\bY  \sim \text{Inverse-Gamma} \left(\frac{n}{2} + \alpha, \frac{\bY^T(\mathbf{I}_n - \bX(\bX^T\bX + \boldsymbol{\Sigma}_{\tau, \omega}^{-1})^{-1}\bX^T)\bY}{2} + \xi \right),\\
& \\
\bbeta | \sigma^2, \btau^2, \bomega^2,\bY \sim {\cal N}_p \left(\left(\bX^T\bX + \boldsymbol{\Sigma}_{\tau, \omega}^{-1}\right)^{-1}\bX^T\bY, \sigma^2 \left(\bX^T\bX + \boldsymbol{\Sigma}_{\tau, \omega}^{-1}\right)^{-1} \right),
\end{cases}
\end{split}
\end{equation}
The one-step transition density of the $\left(\bbeta,  \sigma^2\right)$-chain  is given by
\begin{eqnarray}\label{trans_2bfl}
k_{fl} \left[(\bbeta_0, \sigma^2_0), (\bbeta_1, \sigma^2_1) \right] = \int_{\mathbb{R}^{p}_{+}} \int_{\mathbb{R}^{p-1}_{+}}\pi(\bbeta_1 | \sigma_1^2, \btau^2, \bomega^2, \bY) \pi(\sigma_1^2 | \btau^2, \bomega^2, \bY)  \pi (\btau^2, \bomega^2 | \bbeta_0, \sigma_0^2, \bY)  d\bomega^2 d\btau^2. \hspace{.5cm}
\end{eqnarray}

\blue{

\begin{theorem}\label{boud_bfl}
For any $\alpha, \xi \geq 0$,  the \twoBG chain for the Bayesian fused lasso model is geometrically ergodic. The Markov operator of the  \threeBG chain with transition density $\tilde{k}_{fl}$ is not Hilbert-Schmidt for any $\alpha, \xi \geq 0$. 
\end{theorem}

}

\section{Simulation Studies}
\label{sec:simulation}
We use simulation studies to demonstrate the advantage of the newly developed \twoBG's over existing MCMC computing solutions for the three Bayesian shrinkage models of interest.  Existing methods include \threeBG and HMC. Briefly, HMC is a state of the art generic MCMC technique that has shown remarkable empirical success in exploring many difficult Bayesian posteriors. See \citet{neal:2011} for an introduction. All HMC samplers in our study are run using the Stan software, which simply requires specification of the Bayesian models in Stan language, then the software will automatically derive the log posterior density, and tune and run HMC algorithms accordingly. Note that, similar to that of Gibbs samplers, different representations or parameterizations of the same model will lead to different HMC algorithms. We experimented two ways to implement HMC for the Bayesian group lasso model. The first one uses the conditional prior $\pi(\bbeta | \sigma^2)$ in \eqref{prior_bgl} directly, while the second one uses its hierarchical representation in \eqref{bgl_frame}. For the sparse group lasso and the fused lasso models, we implemented them in Stan where the conditional priors of $\bbeta | \sigma^2$ are directly specified as \eqref{prior_bsgl} and \eqref{prior_bfl}, respectively. Their hierarchical versions are not implemented because they are highly inconvenient to put down in the language of Stan, and are expected to be much more costly per iteration. \blue{The \twoBG and \threeBG samplers are coded in $\text{C}_{++}$, then called from within R using the package \textit{Rcpp}. The code can be found in an online supplement to this paper.}

For the Markov chain underlying each algorithm, let $\rho_k$ denote the lag-$k$ autocorrelation of the $\sigma^2-$marginal of the chain (under stationarity).  \citet[sec~3]{rajar:spar:2015} provided comprehensive reasons why $\rho_1$ is a good proxy for the mixing rate of the joint Markov chain, more so than that of the autocorrelations of the regression coefficients. For the empirical evaluation of an MCMC algorithm, the closer the estimates of $\rho_1$ is to $0$, the better the chain mixes. 
\blue{Another criteria we used to compare the chains is the per unit time effective sample size of the $\sigma^2$ component.} Specifically, $N$ iterations of the Markov chain produce an effective sample size (ESS) of \begin{eqnarray}\label{n_eff}
  	N_\text{eff} = \frac{N}{1+2\sum_{k = 1}^{\infty}{\rho_k}}\,.
  \end{eqnarray}
\blue{We estimate $N_\text{eff}$ of the HMC, the \threeBG and the \twoBG samples using the {\it{effectiveSize}} function from the R package {\it{coda}}~\citep{plummerEtal:2006}.} Then the estimates of $N_\text{eff}$ are divided by $T$, their respective running time in seconds, for the eventual comparisons reported in this section. 
\blue{
\begin{remark} Besides $\rho_1$ and $N_{eff} / T$, we tried to compare the different MCMC algorithms using the multivariate ESS~\citep{wilk:1932, vats:fleg:jone:2019}, carried out using the function {\it{multiESS}} from the R package {\it{mcmcse}}~\citep{FlegalEtal:2017}. Under our simulation setups, the estimates of the multivariate ESS for the full vector of interest $(\sigma^2, \bbeta)$ do not appear stable enough, 
potentially due to the very high dimension of this vector. Hence, we opt to report the estimated multivariate ESS of the 3-dimensional vector $(\sigma^2, \beta_1, \beta_2)$ for our simulation study, which turns out to largely agree with the comparison results based on the (univariate) ESS for the $\sigma^2$ component, $N_\text{eff}$, defined above. The results for multivariate ESS can be found in the supplementary material.
\end{remark}}

\blue{\subsection{Comparing MCMC Algorithms for the Three Bayesian Shrinkage Models}\label{sec:simu3models}}

Data in this study are simulated using the following model:
\begin{eqnarray}\label{simulation_response}
	\bY = \bX \bbeta_* + \boldsymbol{\epsilon},
\end{eqnarray}
where $\boldsymbol{\epsilon}$ is a $n \times 1$ vector of independent standard normal random variables, and $\bbeta_*$ is a $p \times 1$ vector consisting of the true regression coefficients. 


\begin{itemize}

\item {\bf Scenario 1. Grouped explanatory variables.} Denote $\mathcal{K}=\{5, 6, 7, 8, 9, 10, 20, 30, 40, 50\}$. \blue{For $n = 50$, let $K \in \mathcal{K}$. For $n = 100$, let  $K \in 2\mathcal{K}$.  } We generate $K$ variables independently from the multivariate normal distribution with mean vector $0$ and the identity variance-covariance matrix. We build order-5 polynomials based on each variable to get a $n\times p$ design matrix $\bX$, where $p = 5K$ and every 5 consecutive columns belong to a group. For each $p$ we experiment with, $\bbeta_*$ is such that only its first $p/5$ elements are nonzero, and are drawn independently from the $t_2$ distribution.
 
\item {\bf Scenario 2. Similar coefficients for adjacent explanatory variables.}  
 At $n = 100$ and $200$, consider $\frac{p}{n} = \{0.5, 0.6, 0.7, 0.8, 0.9, 1, 2, 3, 4, 5\}$. To generate the $n\times p$ design matrix $\bX$, the rows are independently drawn, each from the $p$-dim multivariate normal distribution, where the mean vector is $0$, and the variance-covariance matrix has $1$ on the diagonal, and $0.2$ everywhere else. Then, the columns of $\bX$ are standardized to have mean zero and squared Euclidean norm $n$. The first and the third $p/10$ elements of $\bbeta_*$ are drawn independently from ${\cal{N}}(1, 0.1^2)$, and all other elements are set to zero. 

\end{itemize}

In analyzing the simulated data, both the Bayesian group lasso and the Bayesian sparse group lasso are applied to scenario~1, and the Bayesian fused lasso to scenario~2. Take the comparison of MCMC algorithms for the Bayesian group lasso model for example. At each $(n, p)$ combination, \blue{$100$} datasets are generated from scenario~1. For every data set, each kind of MCMC algorithm is run for \blue{$10,000$} iterations, with the first \blue{$1,000$} discarded as burn in. Each point in Figures~\ref{fig:bgl},~\ref{fig:bsgl} and~\ref{fig:bfl} represents the average lag-one autocorrelation estimates from \blue{$100$} Markov chains of the same kind, over different datasets. Similarly, each point in Figures~\ref{fig:eff:bgl},~\ref{fig:eff:bsgl} and~\ref{fig:eff:bfl} represents the decadic logarithm of the average of $N_{eff} / T$ over the \blue{$100$}  Markov chains. \blue{To show that the comparison results in the Figures are meaningful and not due to randomness, we include summaries of the standard error of these averages in the main text, and include boxplots for the 100 estimates of $\rho_1$ and that of $N_{eff} / T$ in the supplementary material.}  The series of experiments are run on the High Performance \blue{Argon} Cluster at University of Iowa, on a standard machine with 64 GB RAM, 16 cores and 2.6GHz processor.

\begin{remark} 
\blue{Note that we choose the relatively short chain length of \blue{$10,000$} for each chain when comparing all the methods because the HMC samplers are very costly. We are actually able to easily run each of the \twoBG and \threeBG samplers for $100,000$ iterations in all scenarios. Estimates of $\rho_1$ and $N_\text{eff}/T$ based on the longer chains are only included in the supplementary material, and they are numerically similar to that of the shorter chains reported here.} 
\end{remark}

\begin{remark} The main goal of the our simulation study is to learn the behavior of the different algorithms, not to see which models or hyperparameter values fit each scenario the best. Hence for each dataset, we have only considered the most reasonable model with their default hyperparameter values. We always set $\alpha = \xi = 0$, which lead to improper priors on $\sigma^2$. Also, we set $\lambda = 1$ for the Bayesian group lasso, and $\lambda_1 = \lambda_2 = 1$ for the other two Bayesian models. The resulting posteriors are proper~\citep{kyungEtal:2010,xf:ghosh:2015}. Our future work includes investigating the impact of hyperparameter values through Bayesian cross-validation and Bayesian sensitivity analysis, or doing fully Bayesian analysis that employs hyperpriors. All such endeavors will heavily depend on the fast algorithms developed in this article.
\end{remark}

\subsubsection{Results for the Bayesian Group Lasso}
\label{sec:simu_bgl}
\begin{figure}[h!]
	\includegraphics[width=.5\linewidth]{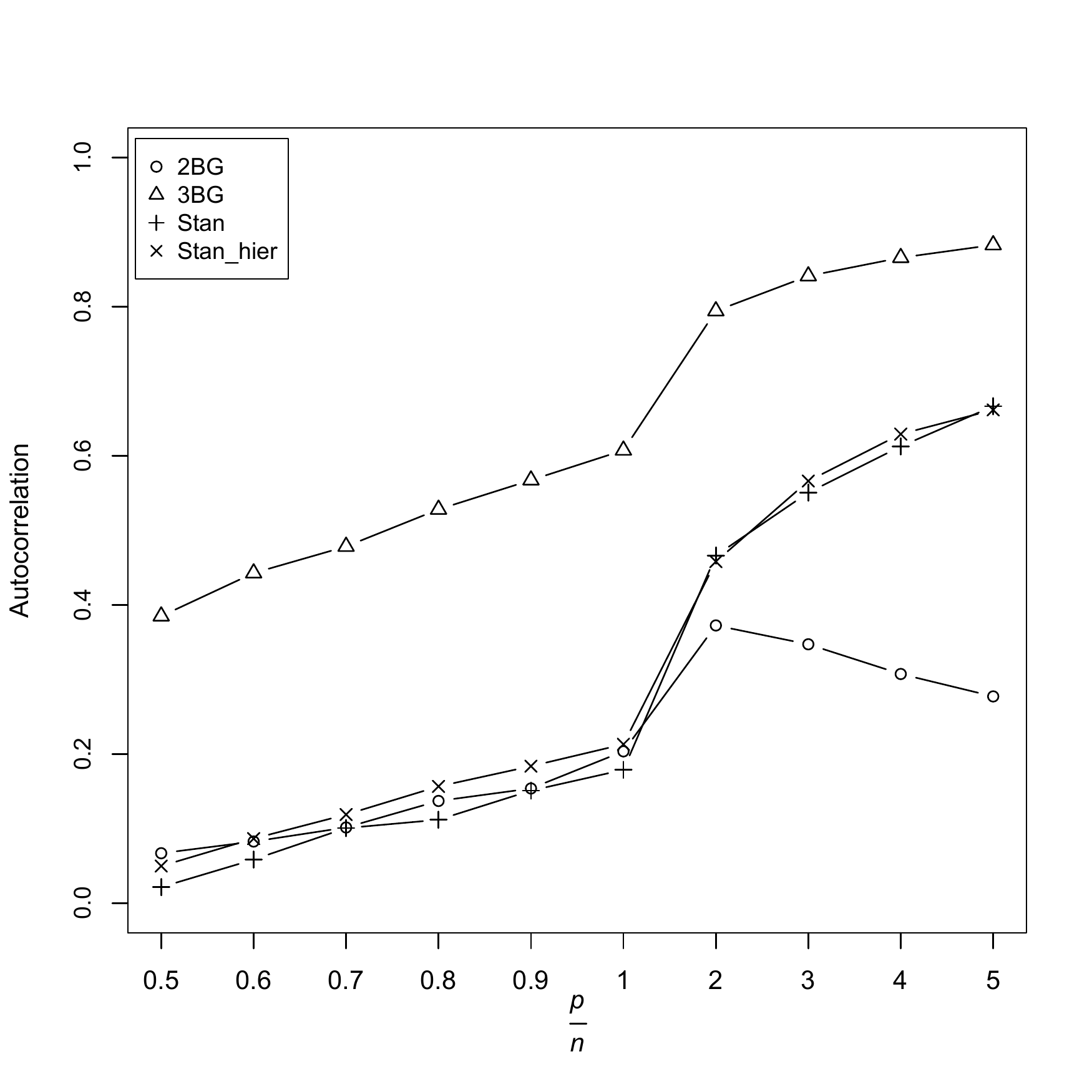}
	\includegraphics[width=.5\linewidth]{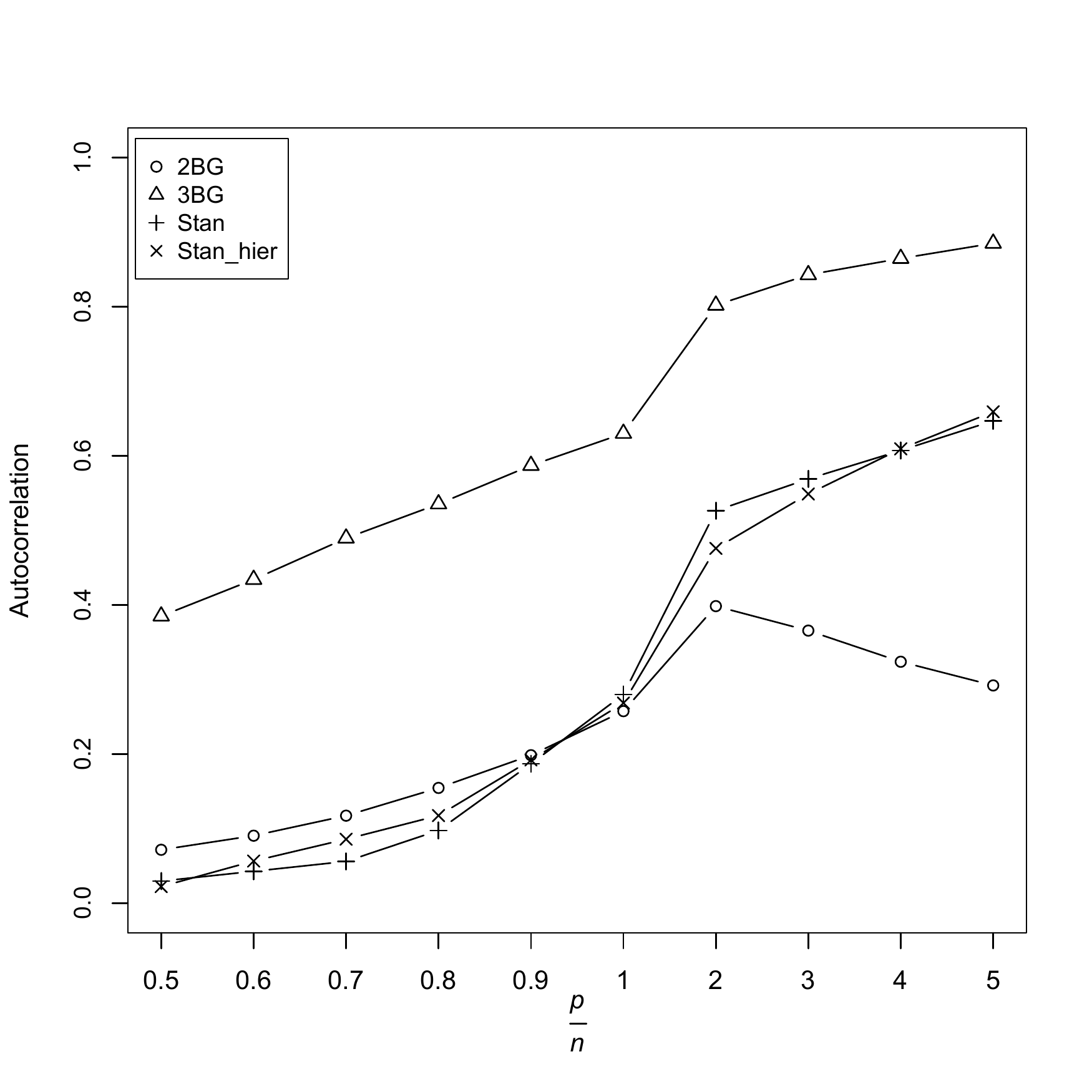}
	\caption{Empirical lag-one autocorrelation of the $\sigma^2$ component of the four MCMC algorithms for the Bayesian group lasso model at $n = 50$ (left) and $n = 100$ (right).}
	\label{fig:bgl}
\end{figure}

\begin{figure}[h!]
	\includegraphics[width=.5\linewidth]{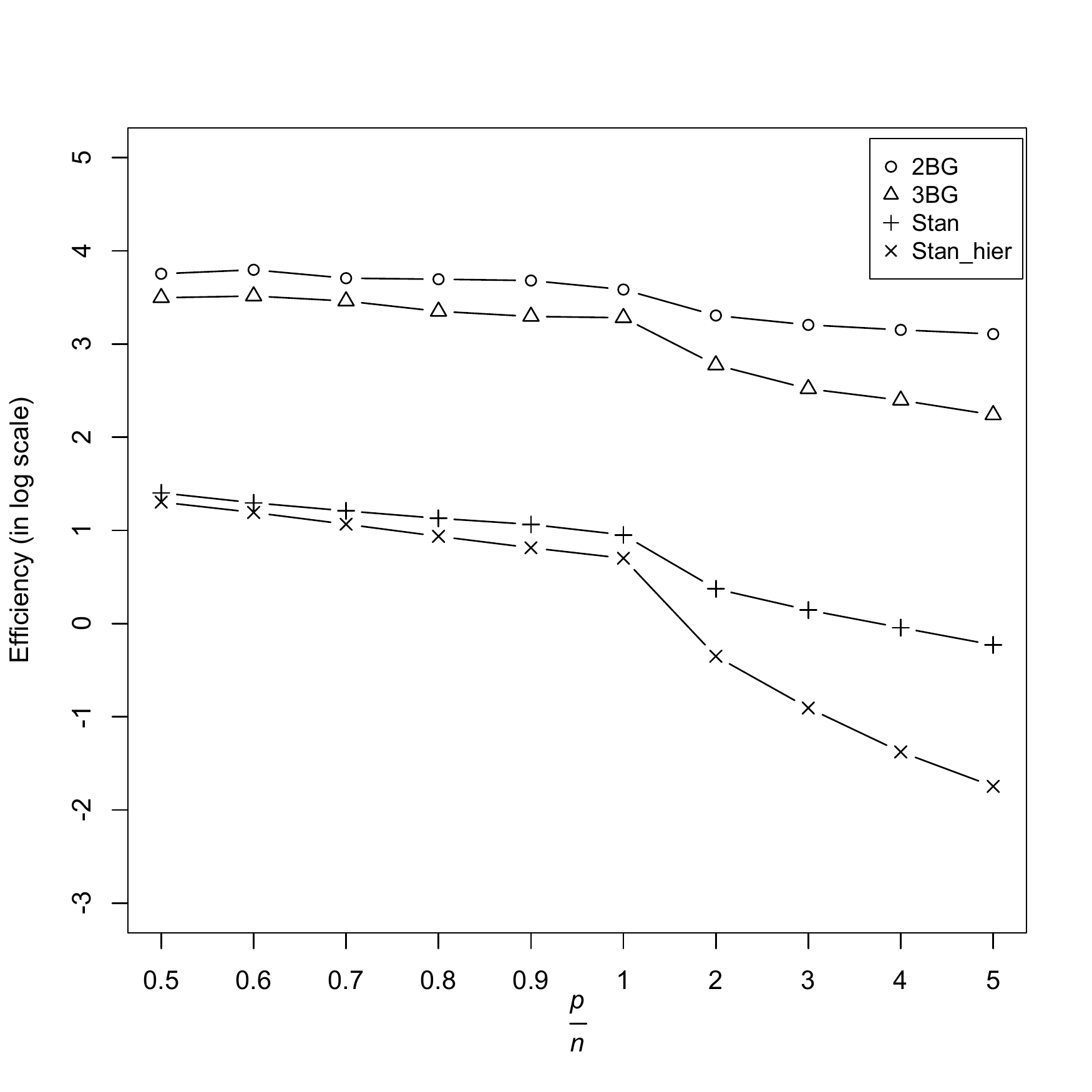}
	\includegraphics[width=.5\linewidth]{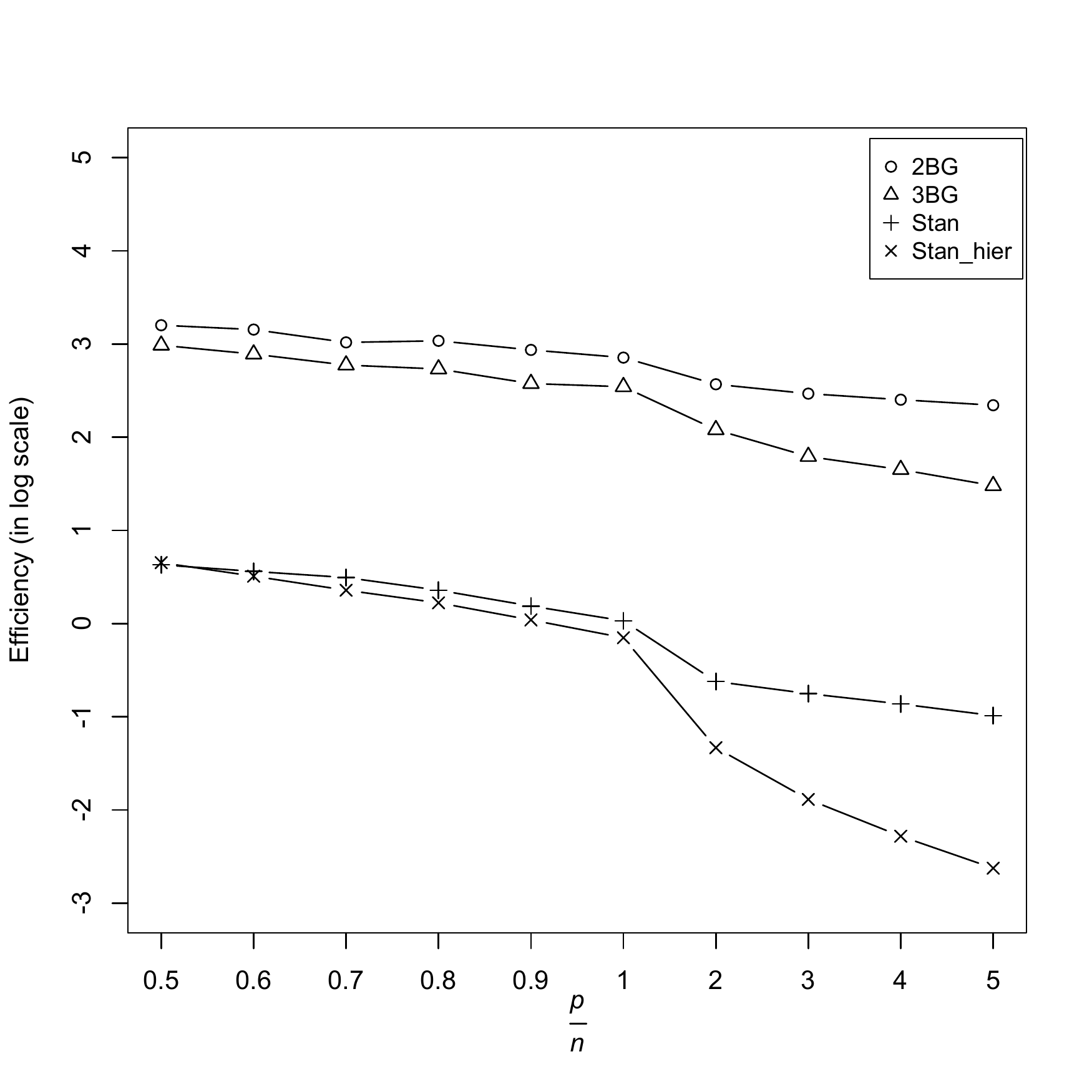}
	\caption{Efficiency, measured by the average effective sample size per second, $N_{eff}/T$, and displayed in base-$10$ log scale, of the four MCMC algorithms for the Bayesian group lasso model at $n = 50$ (left) and $n = 100$ (right).}
	\label{fig:eff:bgl}
\end{figure}

Figure~\ref{fig:bgl} displays the average lag-one autocorrelation of the $\sigma^2$-chain. We can see that the proposed \twoBG has dramatically smaller autocorrelations than the classical \threeBG across all setups of $(n, p)$, indicating better mixing of \twoBG. Both HMC samplers have comparable mixing rates to that of \twoBG when $p$ is no bigger than $n$. But with larger $p$, the lag-one autocorrelation of the HMCs deteriorates, while that of \twoBG stays below $0.4$. \blue{The standard error of the average $\rho_1$ at all setups are calculated, which never exceed $0.0065$ and are typically between $0.00073$ and $0.0064$.} Further, Figure~\ref{fig:eff:bgl} takes computing time into consideration by examining the average effective sample size per second, $N_{eff}/{T}$, in the log scale of each sampler. \blue{The standard error of the average $N_{eff}/{T}$ (in base-$10$ log scale) at all setups are less than $0.021$ with typical values between $0.0047$ and $0.021$.} These observations show that \twoBG is computationally more efficient than all others. For example, at $n = p = 50$, \twoBG produces about twice as many effective samples per second as that of \threeBG, $430$ times that of ``Stan'' and $766$ times that of ``Stan\_hier'. At $n=50, p=250$, the relative efficiency of \twoBG over the other three methods increases to \blue{$7$, $2172$ and $71640$}, respectively. Note that ``Stan\_hier"  is computationally much less efficient than ``Stan", as a result of similar mixing rate of the underlying Markov chains, but that the hierarchical version involves a higher-dimensional posterior that greatly increases the computational cost per iteration. 
 

\subsubsection{Results for the Bayesian Sparse Group Lasso}
\label{sec:simu_bsgl}

\begin{figure}[h!]
	\includegraphics[width=.5\linewidth]{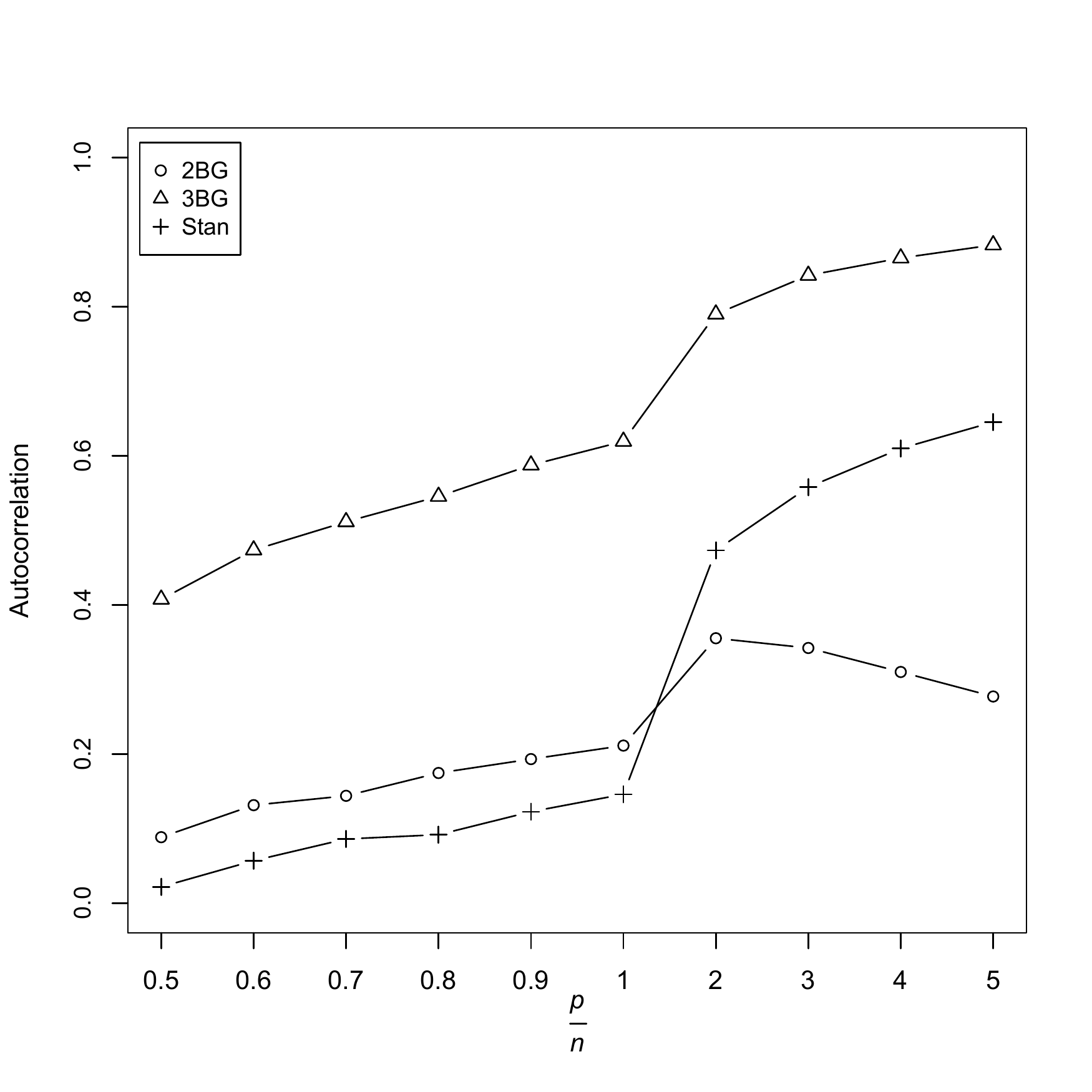}
	\includegraphics[width=.5\linewidth]{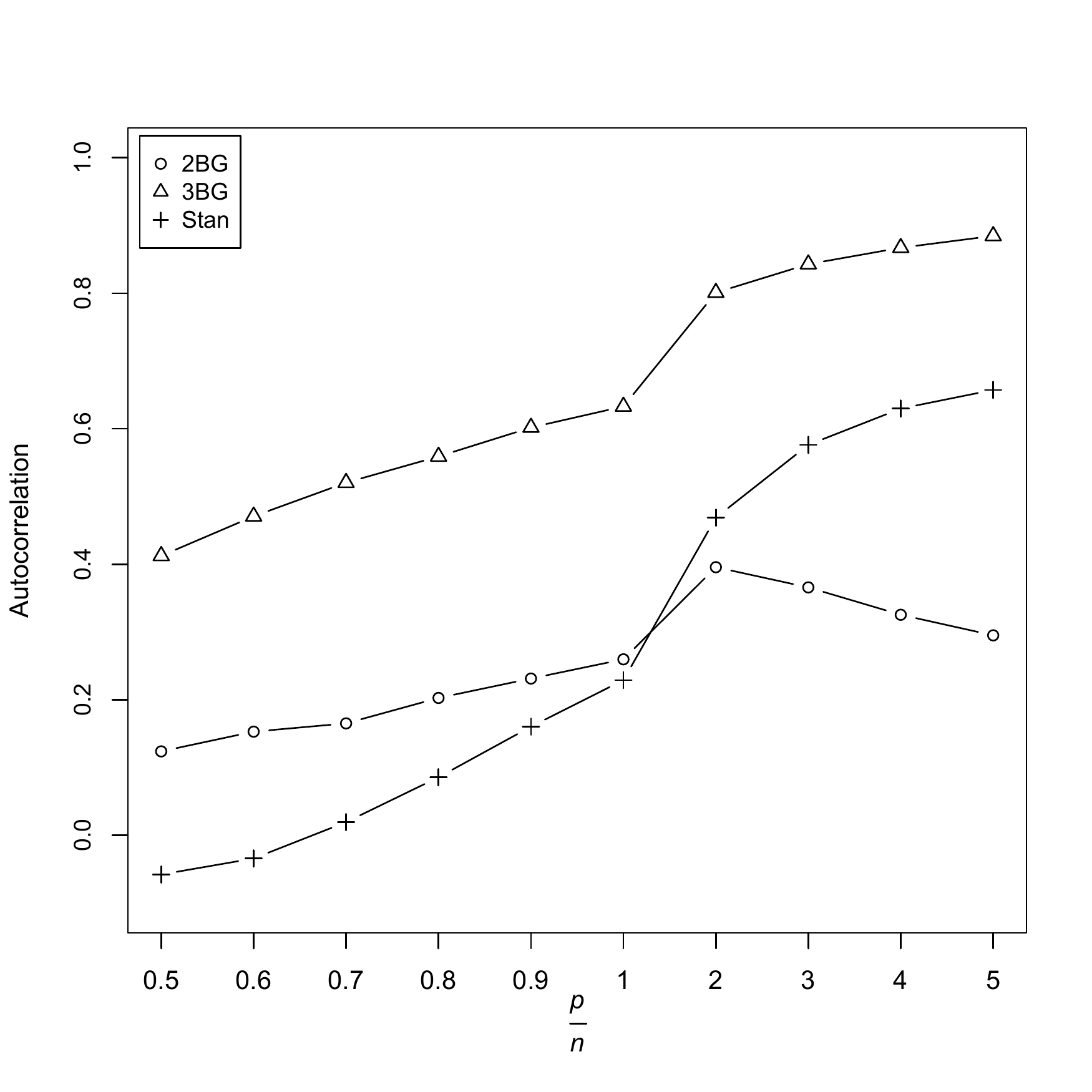}
	\caption{Empirical lag-one autocorrelation of the $\sigma^2$ component of the three MCMC algorithms for the Bayesian sparse group lasso model at $n = 50$ (left) and $n = 100$ (right).}
	\label{fig:bsgl}
\end{figure}
\begin{figure}[h!]
	\includegraphics[width=.5\linewidth]{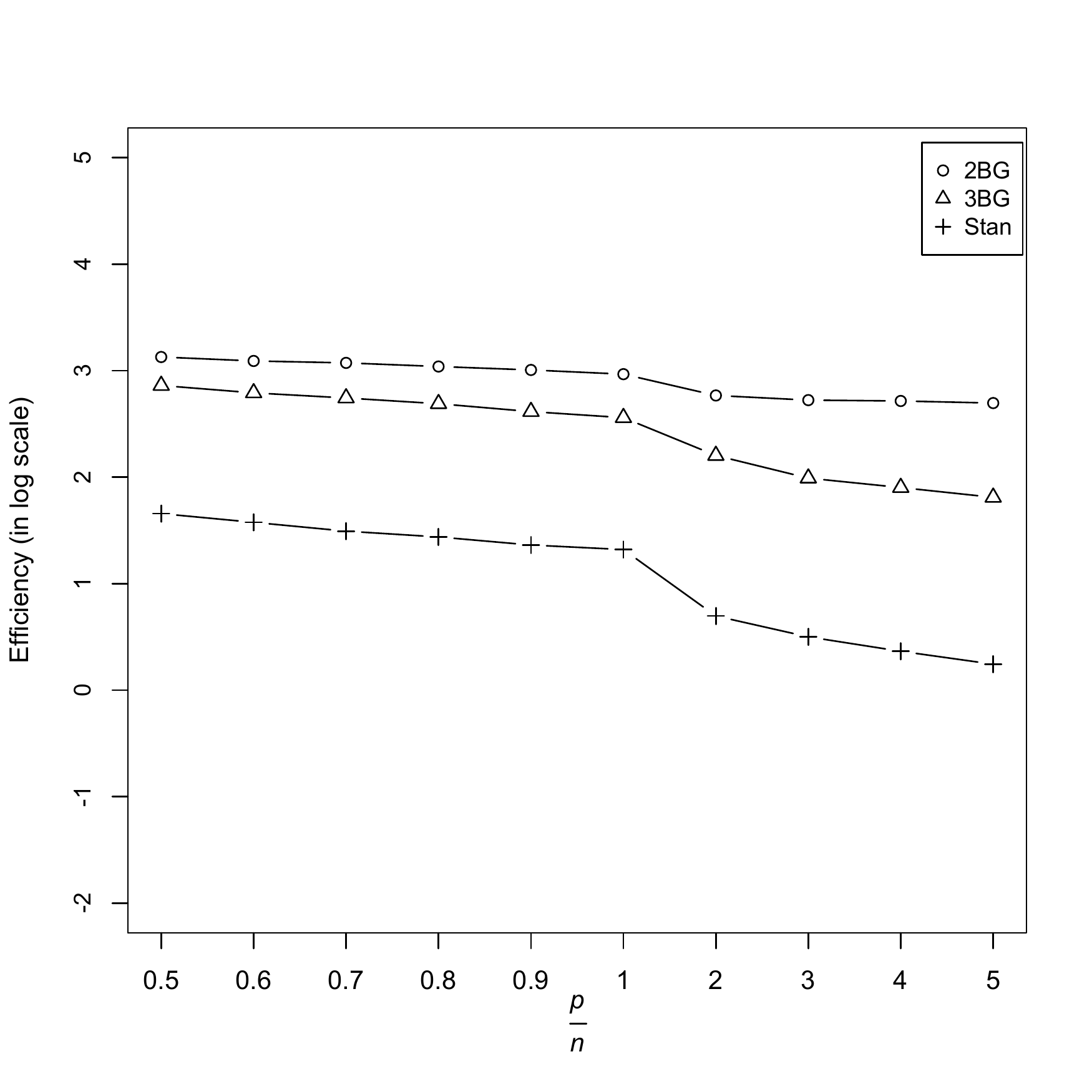}
	\includegraphics[width=.5\linewidth]{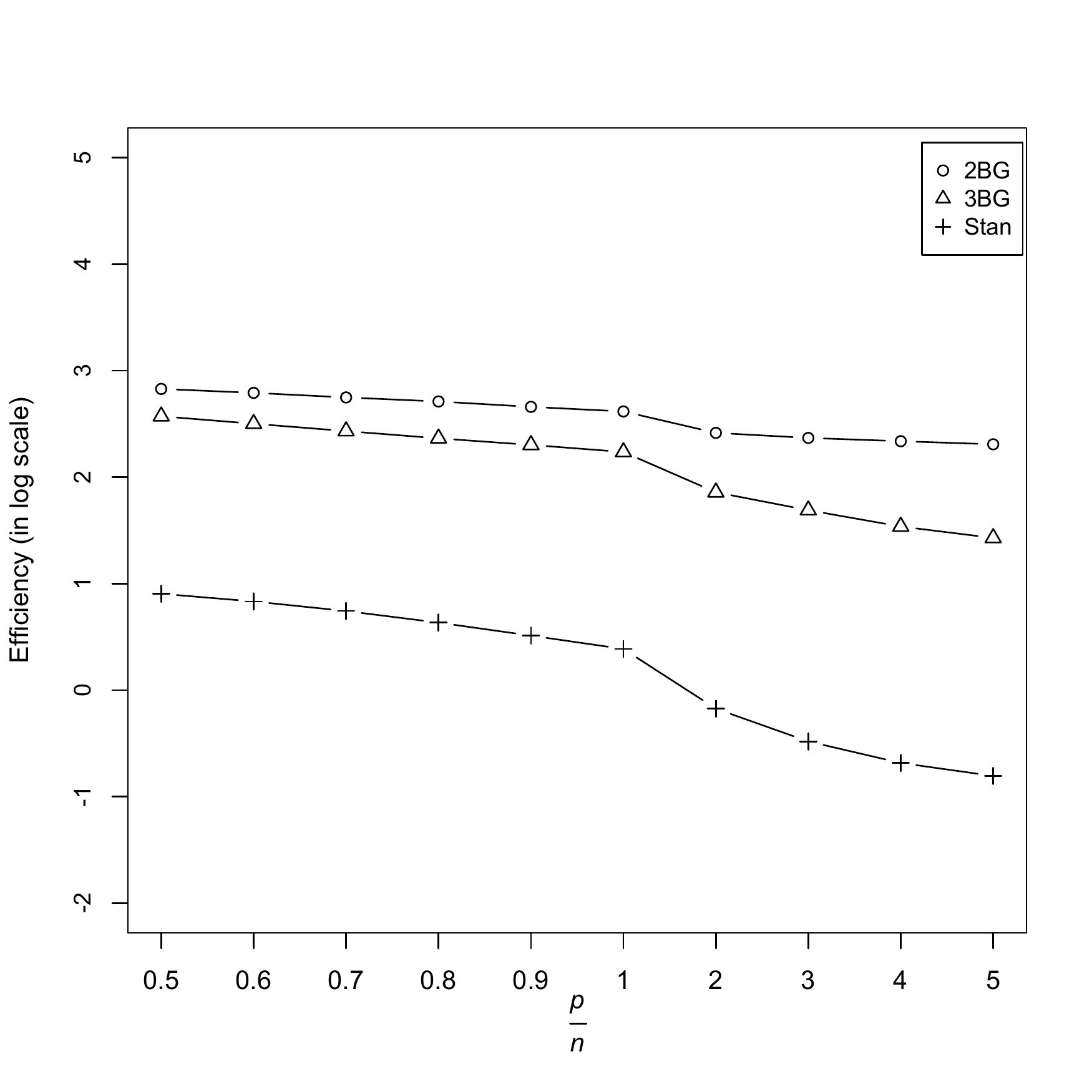}
	\caption{Efficiency, measured by the average effective sample size per second, $N_{eff}/T$, and displayed in base-$10$ log scale, of the three MCMC algorithms for the Bayesian sparse group lasso model at $n = 50$ (left) and $n = 100$ (right).}
	\label{fig:eff:bsgl}
\end{figure}
Figure~\ref{fig:bsgl} displays the lag-one autocorrelation of the $\sigma^2$-chain of the \twoBG, the \threeBG and the HMC sampler for the Bayesian sparse group lasso model, which shows a similar pattern as that of the group lasso model. Again, \twoBG always mixes much faster than \threeBG. HMC mixes faster than \twoBG when $p \leq n $, but \twoBG shows an increasingly large advantage over HMC as $p$ grows larger than $n$. After taking computing time into account, in Figure~\ref{fig:eff:bsgl}, we see that \twoBG produces the most effective samples per second in all setups. For example, at $n = p = 50$, \twoBG produces around $3$ and $43$ times as many effective samples per second as that of \threeBG and HMC, respectively. And at $n=50, p=250$, the relative efficiency of \twoBG to the other two samplers grow to about $9$ and $273$, respectively. \blue{As for the standard error associated with the averages in Figures~\ref{fig:bsgl} and \ref{fig:eff:bsgl}, that of the $\rho_1$ estimates never exceed $0.0087$ with typical values between $0.00081$ and $0.0077$, and that of the $N_{eff}/{T}$ estimates (in base-$10$ log scale) never exceed $0.015$, with typical values between $0.0047$ and $0.014$.  Boxplots of the estimates from all the replications are also available in the supplementary material. These suggest that the patterns observed in Figures~\ref{fig:bsgl} and ~\ref{fig:eff:bsgl}  are meaningful.}

\subsubsection{Results for the Bayesian Fused Lasso}
\label{sec:simu_bfl}
\begin{figure}[h!]
	\includegraphics[width=.5\linewidth]{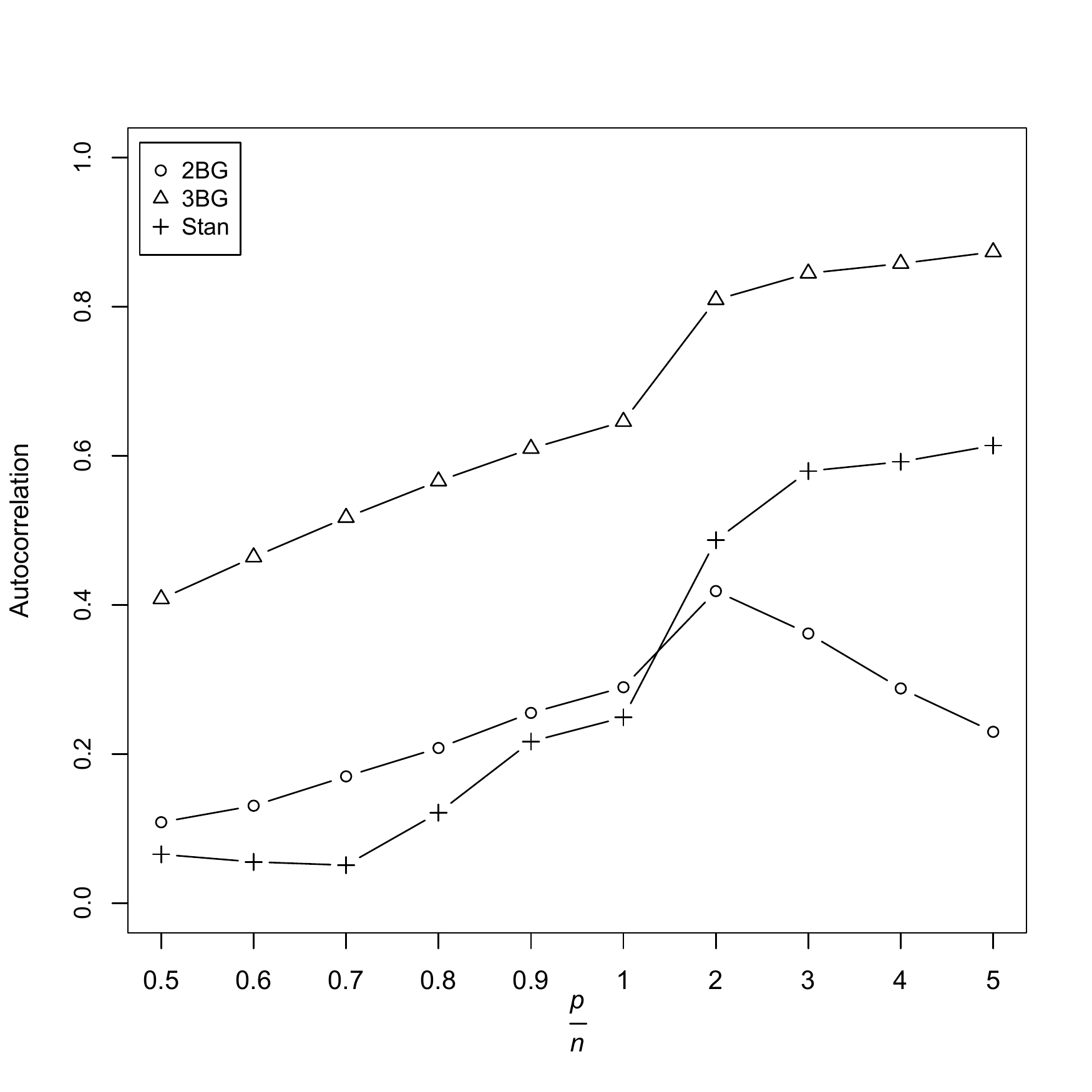}
	\includegraphics[width=.5\linewidth]{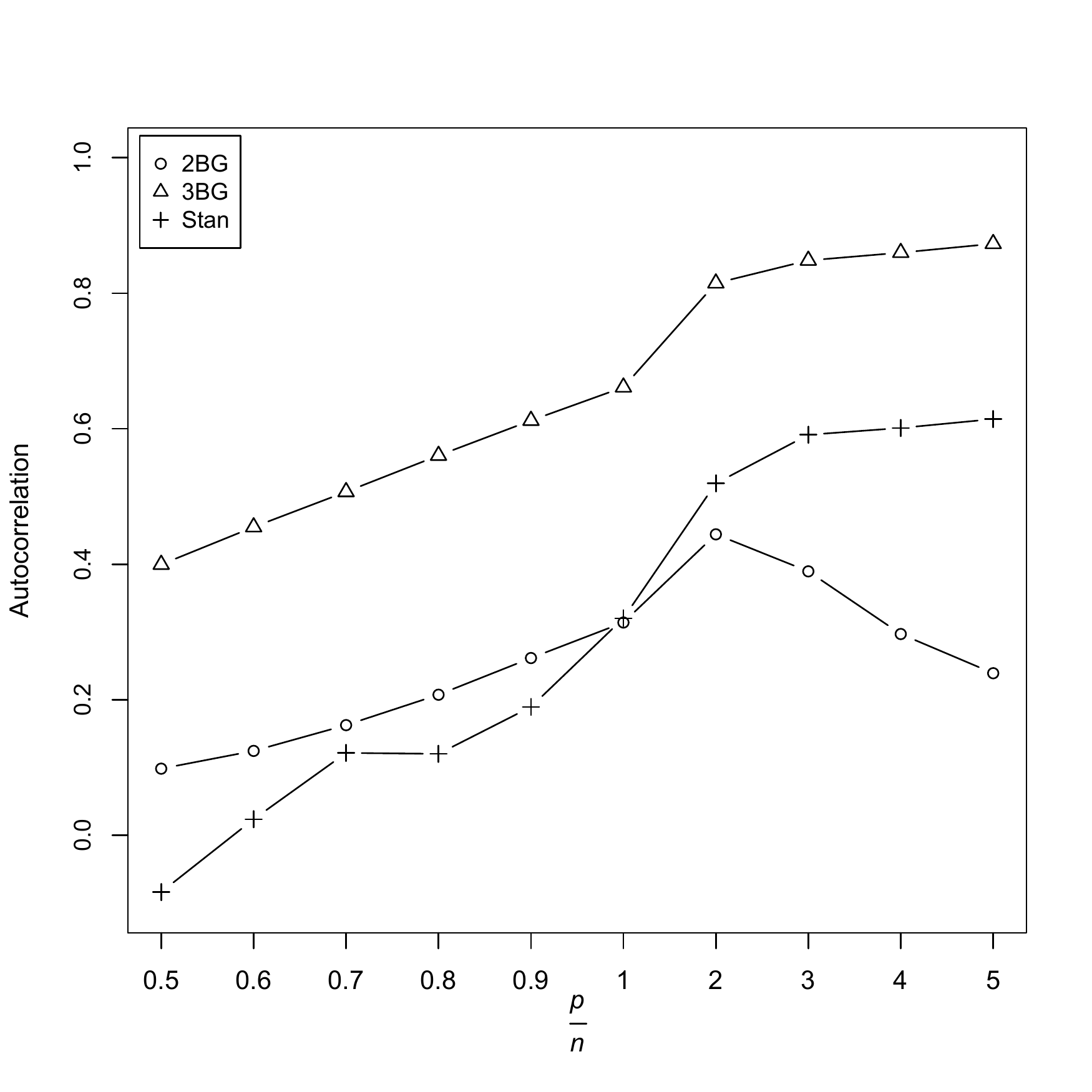}
	\caption{Empirical lag-one autocorrelation of the $\sigma^2$ component of the three MCMC algorithms for the Bayesian fused lasso model at $n = 100$ (left) and $n = 200$ (right).}
	\label{fig:bfl}
\end{figure}

\begin{figure}[h!]
	\includegraphics[width=.5\linewidth]{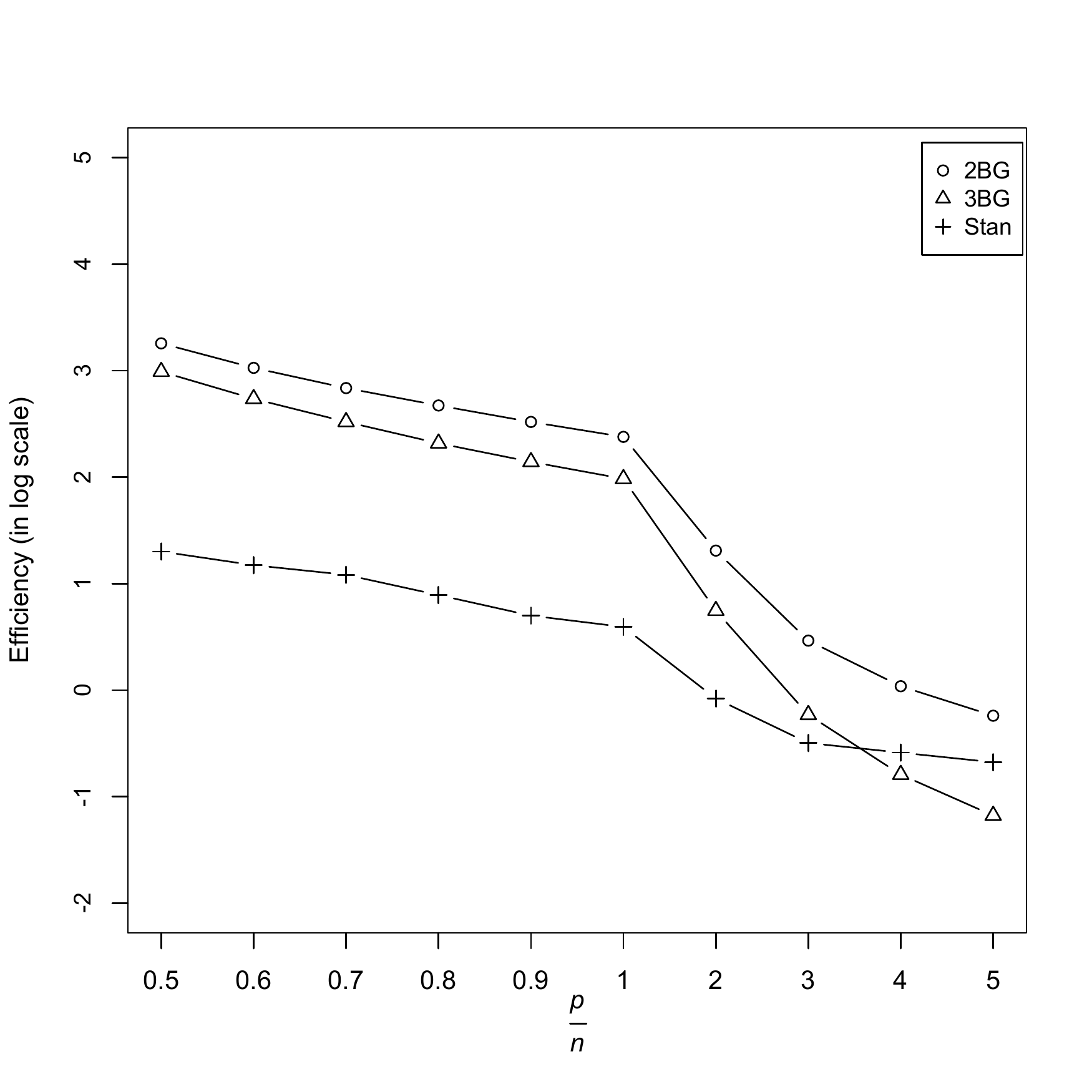}
	\includegraphics[width=.5\linewidth]{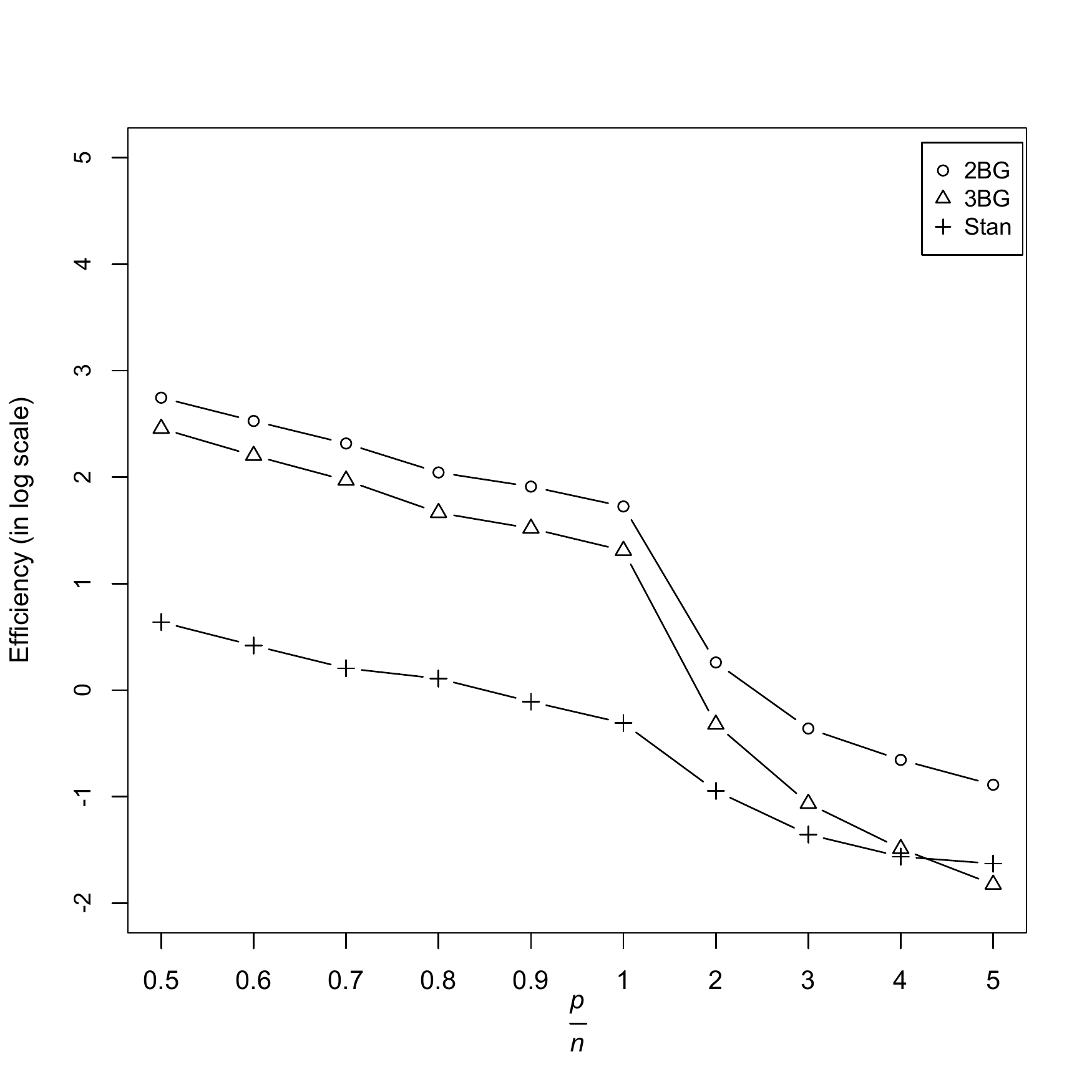}
	\caption{Efficiency, measured by the average effective sample size per second, $N_{eff}/T$, and displayed in base-$10$ log scale, of the three MCMC algorithms for the Bayesian fused lasso model at $n = 100$ (left) and $n = 200$ (right). }
	\label{fig:eff:bfl}
\end{figure}

Figure~\ref{fig:bfl} displays the autocorrelation of the $\sigma^2$-chain of the three samplers for the Bayesian fused lasso model. \twoBG and HMC always mix much faster than \threeBG, while \twoBG has an advantage over HMC for large values when $p$ exceeds $n$. Taking computing time into account, Figure~\ref{fig:eff:bfl} shows that \twoBG is the most efficient sampler in all situations. For example, at $n = p= 100$, the \twoBG produces $3$ times as many effective samples as that of \threeBG and $60$ times that of ``Stan''. When $p$ increases to $500$, \twoBG is $8$ times as efficient as \threeBG and twice that of HMC, though all three generate less than $1$ effective sample per second. \blue{The standard error of the averages are small, that for the $\rho_1$ estimates never exceed $0.0046$ and are typically between $0.00058$ and $0.0044$, and that for the $N_{eff}/{T}$ estimates (in the log scale) never exceed $0.018$ and are typically between $0.0032$ and $0.018$.  Boxplots of the estimates from all the replications are also available in the supplementary material. These suggest that the patterns observed in Figures~\ref{fig:bfl} and ~\ref{fig:eff:bfl}  are meaningful.}

\subsection{\blue{Analyzing Extra Wide and Extra Tall Datasets with Bayesian Group Lasso Models} }
\label{sec:simu_add}

\blue{We perform additional studies following directions suggested by a reviewer, including analyzing datasets that are extra wide, extra tall and have non-normal noises. The first two are presented below, and the third in the supplement.}  

\begin{figure}[h!]
	\includegraphics[width=.5\linewidth]{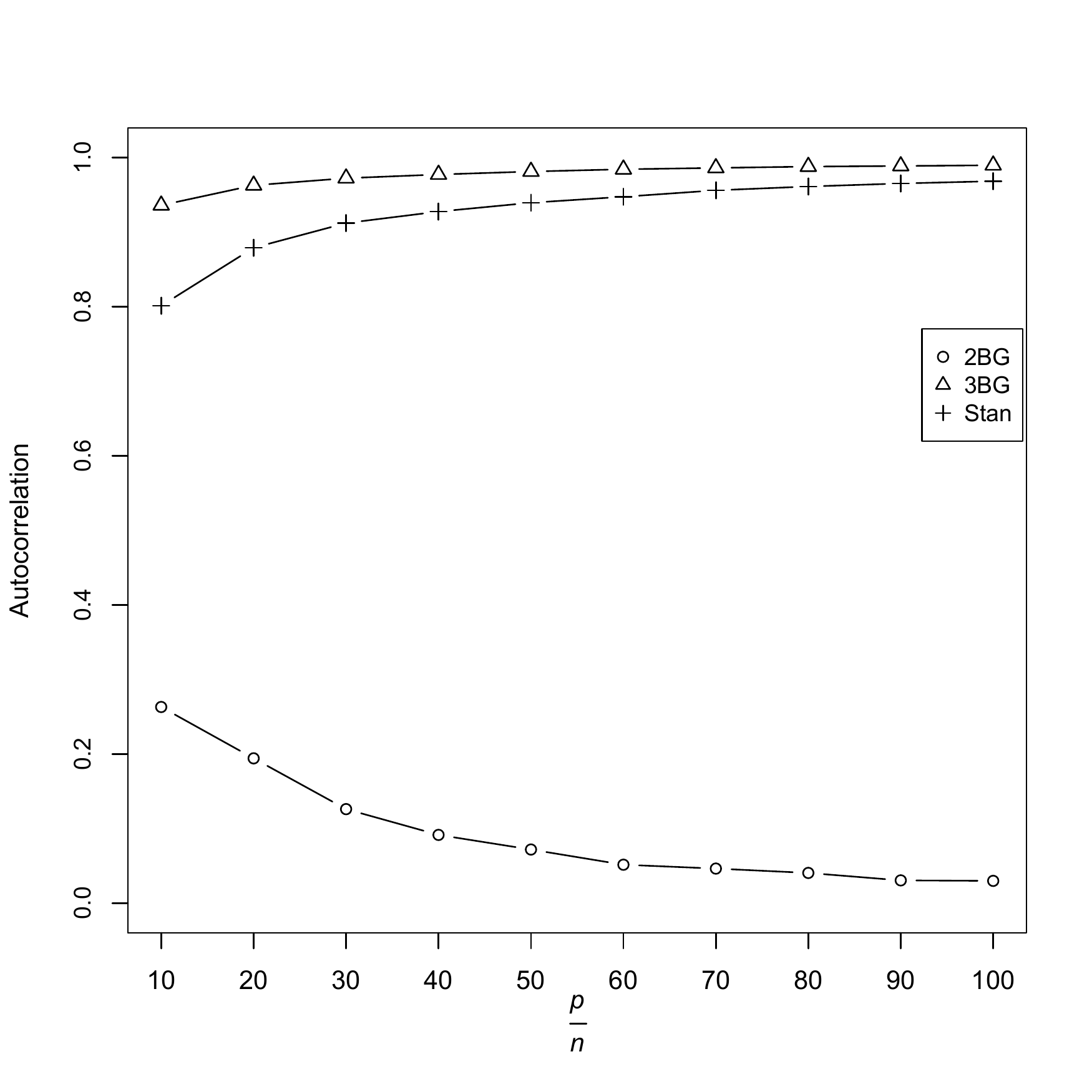}
	\includegraphics[width=.5\linewidth]{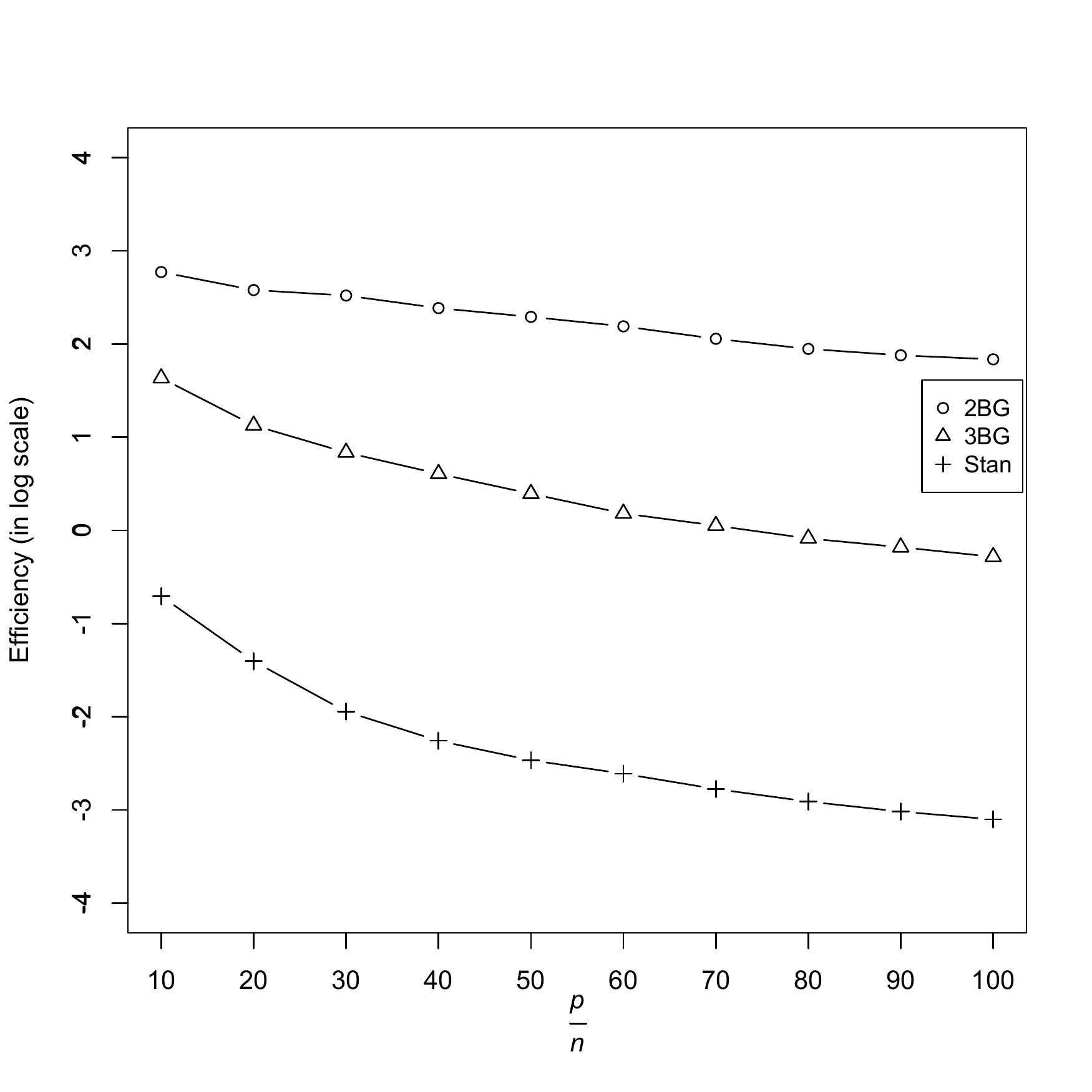}
	\caption{Empirical lag-one autocorrelation (left) and estimated effective sample size per second (right) of the $\sigma^2$ component of the three MCMC algorithms for Bayesian group lasso models with extra wide datasets.  }
	\label{fig:imba}
\end{figure}

\blue{First, we consider extra wide datasets and restrict the study to Bayesian group lasso models due to space limit. We form datasets with $n = 50$ and $\frac{p}{n} = (10, 20, 30, \cdots, 100)$, respectively. For each $p$, the design matrix $\boldsymbol{X}$ is generated using the same scheme as that of scenario~1 in section~\ref{sec:simu3models} and $\bbeta_*$ is such that only its first $5$ elements are nonzero, and are drawn independently from the $t_2$ distribution. Note that we fix the number of true covariates to achieve more and more sparse signals as $p$ increases, which forms more extreme setups than the ones in section~\ref{sec:simu3models}. At each $(n, p)$ combination, $100$ datasets are generated, and for every data set, each kind of MCMC algorithm is run for \blue{$10,000$} iterations, with the first \blue{$1,000$} discarded as burn in. 
Figure~\ref{fig:imba} displays the average lag-one autocorrelations and the average estimates of $N_{eff} / T$ for the $\sigma^2$-chain, and corresponding boxplots can be found in the supplement. We see that, as $p$ grows much larger than $n$, the mixing rates of \threeBG and HMC deteriorate, while that of the \twoBG actually improves.}

\begin{figure}[h!]
	\includegraphics[width=.5\linewidth]{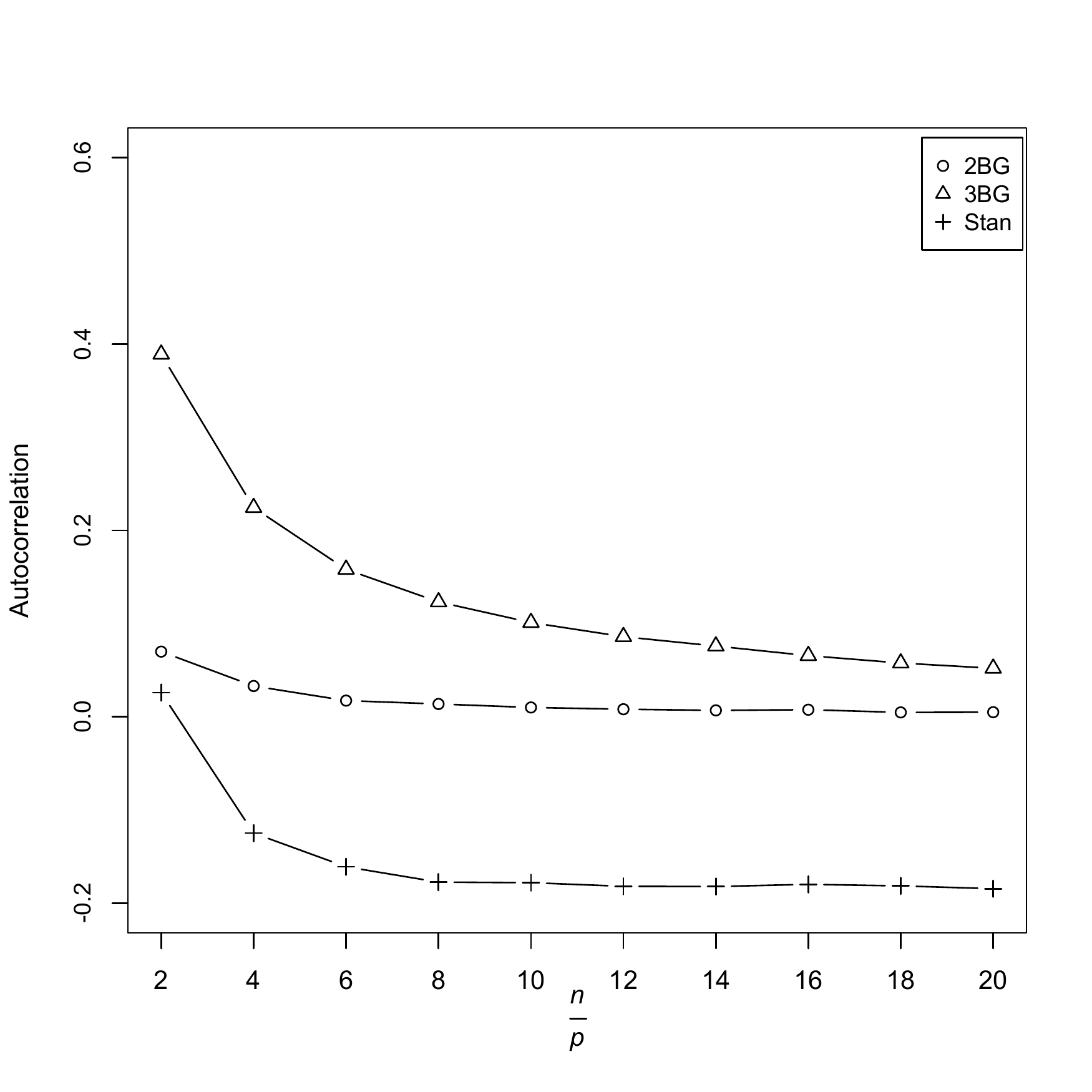}
	\includegraphics[width=.5\linewidth]{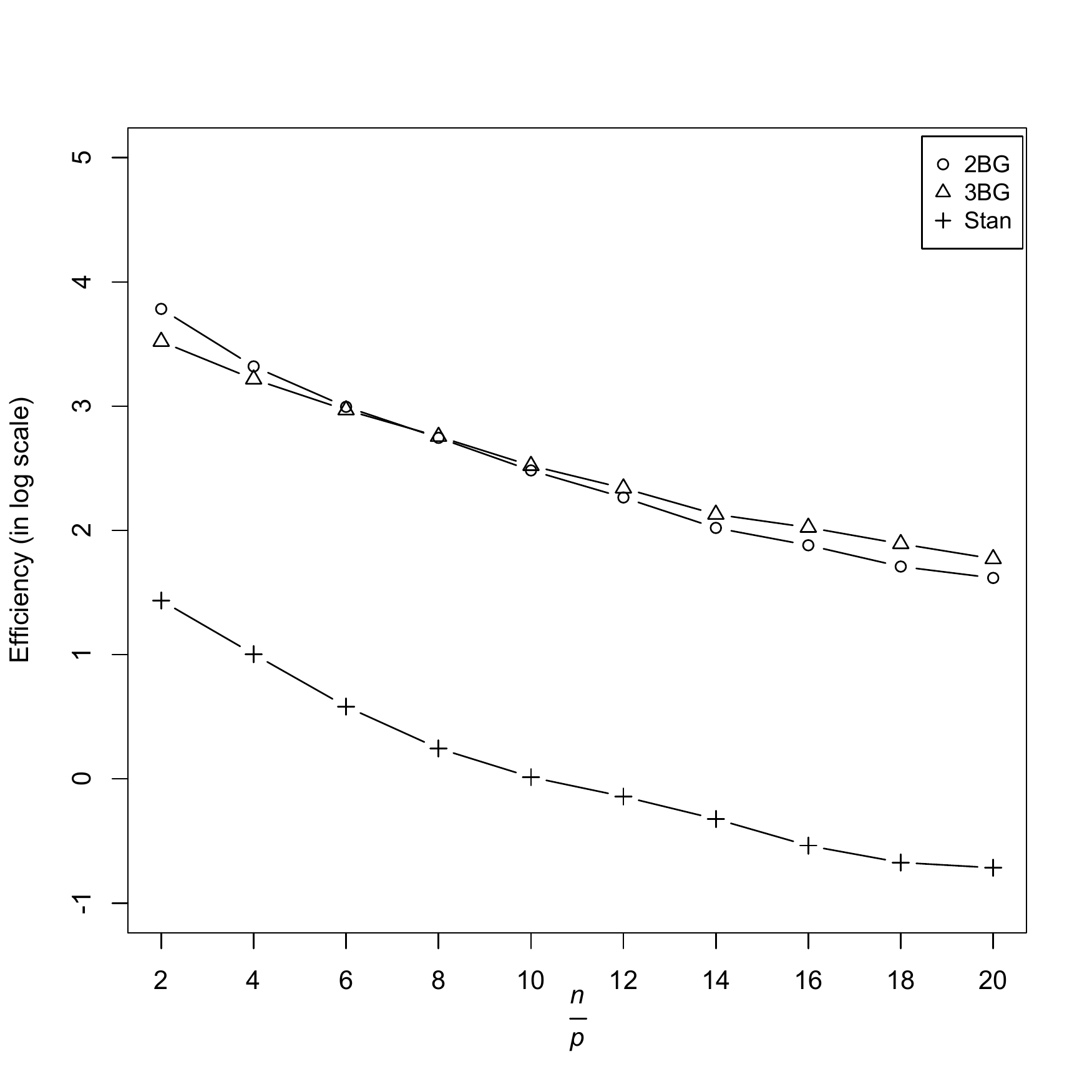}
	\caption{ Empirical lag-one autocorrelation (left) and estimated effective sample size per second (right) of the $\sigma^2$ component of the three MCMC algorithms for Bayesian group lasso models with extra tall datasets.  }
	\label{fig:tall}
\end{figure}

\blue{Next, we consider Bayesian group lasso models applied to tall datasets. Specifically, for $p = 25$ and $n \in \{50, 100, 150, \cdots, 500\}$, $100$ datasets are generated for each $(n, p)$ combination using the same generation scheme as above. 
MCMC samples of the three algorithms are generated in the same manner as before. The left panel of Figure~\ref{fig:tall} shows that all Markov chains mix quite well for tall datasets, where HMC mixes the best and often enjoys negative autocorrelations, and \twoBG generates almost independent samples. The right panel of Figure~\ref{fig:tall} shows that, once taking computing time into consideration, \twoBG is still more efficient than HMC. It's also interesting to see that the ESS per second of \threeBG is comparable to that of \twoBG. Indeed, despite having slightly worse mixing rates at large $n$, \threeBG chains cost a little less per iteration than \twoBG in these setups.}


\subsection{Summary of Simulation Results}
\label{sec:simu_summary}

For all three Bayesian shrinkage models,  \twoBG is almost always the most efficient among existing algorithms. 
In the most demanding, high-dimensional regression problems where the number of predictors $p$ is larger than the data size $n$, \twoBG produces about $10$ -- $10^4$ times as many effective samples per second as that of its competitors. Contributing to the advantage of \twoBG are (1) the fast mixing rate of \twoBG, which is similar to that of the state of the art method HMC for small $p$, and even faster than that of the HMC for larger $p$, and (2) the low computing cost per iteration of \twoBG, which is similar to that of \threeBG, and much lower than that of HMC.

\blue{ 
Besides the theoretical results on \twoBG and \threeBG chains presented in section~\ref{sec:fast}, our intuition for why \twoBG mixes much better than HMC and \threeBG in our studies are as follows. For Bayesian shrinkage models, \cite{rajar:spar:2015} established that $\bbeta$ and $\sigma^2$ has high posteriori dependence when $p >> n$. For HMC in general, \cite{sam:giac:2019} argued that HMC is highly sensitive to tuning when the posterior features strong dependence among components, partly because the spectral gap of its Markov operator decays to zero very fast when the mismatch between the proposal and target scales increases in some directions. Hence, for Bayesian shrinkage models with large $p$, HMC can be extremely hard to tune to overcome the high posterior dependence given the current techniques and softwares, as we see in Figure~\ref{fig:imba} for the example of extra wide datasets. Similarly, \threeBG involves drawing from the conditionals $\bbeta| \sigma^2, \tau$ and $\sigma^2|\bbeta, \tau$, which produces high auto-correlations between successive $\bbeta$ and $\sigma^2$  samples in the presence of high posteriori dependency between these two components. In contrast, the newly proposed \twoBG avoid the high dependence problem by updating $\bbeta$ and $\sigma^2$ within the same block, and hence explore the posterior distribution of high-dimensional Bayeisan shrinkage models more efficiently. \blue{See \citet{junEtal:1994} for some general theory on the benefit of blocking in Markov chains.}} 

\section{Real Data Analysis}
\label{sec:realdata}
In this section, we observe superior performance of the proposed \twoBG over \threeBG and HMC in real data applications. 
Three data sets are analyzed, each using a different Bayesian shrinkage model that is appropriate for the structure of the covariates. For each model, all three computing methods are used and evaluated in terms of mixing rate and computational efficiency, as summarized in Table~\ref{tab:summary}. 

Each algorithm is run for $20,000$ iterations with the first one-tenth discarded as ``burn-in".  The experiments are run on a 2.2 GHz machine with 16GB RAM and MacOS Sierra system. More details of the data and the comparisons in computation are given in the subsections below.

\begin{table}[h]
	\begin{center}
		\begin{tabular}{ccccccccccc}
			\hline
		 &  &  & \mc{3}{Autocorrelation} & & \mc{4}{$ \frac{N_{eff}}{T} $} \\	
		Data Set & n & p & \twoBG & \threeBG & Stan & & \twoBG & \threeBG & Stan & \\ \hline 
		Gene Expression Data & 120 & 100 & \bf{0.057} & 0.40&  0.19&  & \bf{578} &  245 &0.08 &  \\
		Economic Data & 72 & 51 & \bf{0.014}& 0.42& 0.029 & & \bf{1213} & 494 & 8.63 & \\
		CGH Data & 200 & 200 & 0.288 & 0.637& \bf{0.232}& & \bf{49} & 19 & 1.38& \\
		     \hline
	\end{tabular}	
		\caption{Lag-one Autocorrelation and effective sample size per second for the $\sigma^2$-component of \twoBG,  \threeBG and HMC as applied to the three datasets in section~\ref{sec:realdata}. }
		\label{tab:summary}
	\end{center}
\end{table}

\subsection{Gene Expression Data}
\label{sec:eye}
We first evaluate the performance of \twoBG, \threeBG and HMC for a Bayesian group lasso model applied to a gene expression data of~\cite{scheetzEtal:2006}, which is publicly available as the data set 
{\it{bardet}} in the {\it{R}} package {\it{gglasso}}~\citep{yang:zou:2015}. The response variable is the expression level of gene {\bf{TRIM32}}, which causes Bardet-Biedl syndrome. The expression level of $20$ other genes are available and expanded using $5$ basis B-splines. The resulting design matrix $\bX$ contains $p=100$ columns where each consecutive $5$ correspond to the same gene. Each column is then standardized to have mean zero and squared Euclidean norm $n$. The Bayesian group lasso model was implemented with regularization parameter $\lambda = 0.06$, based on a cross-validation step of its frequentist counterpart. Details can be found in the supplementary materials.

The \twoBG chain mixes very fast with an average lag-one autocorrelation of $0.057$ for the $\sigma^2$-chain, compared to that of $0.19$ of HMC, and $ 0.40$ of \threeBG. Next, consider the number of effective samples of the $\sigma^2$-component generated per second, \twoBG comes first at around $579$, followed by \threeBG at $245$, and HMC at only $0.08$.

\subsection{Economic Data }
\label{sec:econ}
We use the Bayesian sparse group lasso model to analyze an economic data from~\cite{rose:spiegel:2010}, which is publicly available as the data set {\it{Crisis2008BalancedData}} in the {\it{R}} package {\it{BSGS}}~\citep{lee:chen:2015}.  The response variable is the growth rate of a crisis measure from 2008 to 2009. The design matrix $\bX$ contains $51$ explanatory variables for the 2008 financial crisis for $72$ countries.  These $51$  variables can be classified into $9$ groups. All the columns of $\bX$ are standardized except for those corresponding to the dummy variables. The regularization parameters were taken to be $\lambda_1 = 0.104$ and $ \lambda_2 = 0.082$. 

Again, using the lag-one autocorrelation of the chain $\sigma^2$, \twoBG comes in first at $0.014$, followed by HMC at $0.029$ and \threeBG at a substantially larger correlation at $0.42$. In terms of the number of effective samples per second, $N_{eff}/T$, \twoBG is the best at $1,213$, followed by $494$ of \threeBG, and $9$ of HMC.

\subsection{Comparative Genomic Hybridization (CGH) Data}
\label{sec:cgh}
We apply the Bayesian fused lasso model to a subset of the CGH data of~\cite{tibshirani:pei:2007}, which is publicly available as the data set {\it{CGH}} in the {\it{R}} package {\it{cghFLasso}}. \cite{tibshirani:pei:2007} analyzed this data with fused lasso model. The response contains $200$ samples of CGH array. The design matrix is the identity matrix. The regularization parameters were taken to be $\lambda_1 =0.129 $ and $\lambda_2 = 0.962$. 

Concerning the lag-one autocorrelation of the $\sigma^2$-chain, \twoBG and the HMC are comparable, at $0.29$ and $0.23$, respectively. While \threeBG falls behind at $0.64$. After taking time into account, \twoBG is clearly the winner by producing $48.86$ effective samples per second, compared to $19.14$ from \threeBG, and $1.38$ from HMC.

\section{Discussion}

As we have seen above, various heuristic arguments and theoretical results suggest that \twoBG is expected to be more efficient than \threeBG, and the empirical evidence in sections~\ref{sec:simulation} and \ref{sec:realdata} show that the improvements are substantial. There are currently no theoretical result that directly compares \twoBG, which is a simple Gibbs type algorithm, to HMC, which is a gradient based algorithm. It may be surprising to many that iteration-wise, \twoBG mixes better than HMC in the most challenging high-dimensional cases, while performing similarly in other cases. And the advantage of \twoBG is more striking when computing cost is taken into account. For example, to obtain $10,000$ effective samples for the Bayesian group lasso model with a data of size $p=5n=500$, it roughly takes \twoBG 1 minute, \threeBG 10 minutes, and HMC (via stan) 13.5 hours of computing time. That is, when using these Bayesian shrinkage models in some moderate to high-dimensional setups, \twoBG could be the only viable MCMC solution.

Another interesting observation common to all experiments in section~\ref{sec:simulation} is that, while the autocorrelations of \threeBG and HMC chains increase quickly towards $1$ as  $p/n$ increases, that of \twoBG chains stay bounded and actually decrease. One of our on-going work is performing complexity analysis of the \twoBG algorithms, that is, studying their asymptotic geometric convergence rate as the dimension of the data increases. Although such theoretical results are extremely hard to obtain, and the proofs need to be tailored to each model specification, there have been several successful examples along with new analysis methods. See, e.g., \citet{qin:hobe:2018, qin:hobe:2019, yang:rose:2017}.

An important practical issue of applying any Bayesian models for data analysis is assigning priors or specifying values for the hyperparameters. This topic is beyond the scope of the current paper but deserves further research. Bayesian cross-validation and Bayesian sensitivity analysis provide ways to calibrate and evaluate different choices of priors and hyperparameters. The fast algorithms developed here serve as critical building blocks to perform these analysis that is computationally more demanding.

\section{Supplementary Materials}
\begin{description}

\item[Additional Results] A document containing background knowledge and proofs of theoretical results, details on the application of the models, and additional simulation results. (pdf)

\item[R code] Code for the examples in this paper. (zip)

\end{description}


\begin{thebibliography}{64}
\expandafter\ifx\csname natexlab\endcsname\relax\def\natexlab#1{#1}\fi
\expandafter\ifx\csname url\endcsname\relax
  \def\url#1{\texttt{#1}}\fi
\expandafter\ifx\csname urlprefix\endcsname\relax\def\urlprefix{URL }\fi

\bibitem[{Armagan et~al.(2013)Armagan, Dunson and Lee}]{arma:duns:lee:2013}
\text{Armagan, A.}, \text{Dunson, D.~B.} and \text{Lee, J.} (2013).
\newblock {Generalized double Pareto shrinkage}.
\newblock \textit{Statistica Sinica} \textbf{23} 119.

\bibitem[{Bhattacharya et~al.(2015)Bhattacharya, Pati, Pillai and
  Dunson}]{bhat:pati:pill:duns:2015}
\text{Bhattacharya, A.}, \text{Pati, D.}, \text{Pillai, N.~S.} and
  \text{Dunson, D.~B.} (2015).
\newblock {Dirichlet--Laplace priors for optimal shrinkage}.
\newblock \textit{Journal of the American Statistical Association} \textbf{110}
  1479--1490.

\bibitem[{Carpenter et~al.(2017)Carpenter, Gelman, Hoffman, Lee, Goodrich,
  Betancourt, Brubaker, Guo, Li and Riddell}]{carpenterEtal:2017}
\text{Carpenter, B.}, \text{Gelman, A.}, \text{Hoffman, M.~D.}, \text{Lee, D.},
  \text{Goodrich, B.}, \text{Betancourt, M.}, \text{Brubaker, M.}, \text{Guo,
  J.}, \text{Li, P.} and \text{Riddell, A.} (2017).
\newblock {Stan: A probabilistic programming language}.
\newblock \textit{Journal of statistical software} \textbf{76}.

\bibitem[{Carvalho et~al.(2010)Carvalho, Polson and
  Scott}]{carv:pols:scot:2010}
\text{Carvalho, C.~M.}, \text{Polson, N.~G.} and \text{Scott, J.~G.} (2010).
\newblock The horseshoe estimator for sparse signals.
\newblock \textit{Biometrika} \textbf{97} 465--480.

\bibitem[{Castillo et~al.(2015)Castillo, Schmidt-Hieber and Van~der
  Vaart}]{cast:schm:vand:2015}
\text{Castillo, I.}, \text{Schmidt-Hieber, J.} and \text{Van~der Vaart, A.}
  (2015).
\newblock Bayesian linear regression with sparse priors.
\newblock \textit{The Annals of Statistics} \textbf{43} 1986--2018.

\bibitem[{Chan and Geyer(1994)}]{chan:geye:1994}
\text{Chan, K.~S.} and \text{Geyer, C.~J.} (1994).
\newblock Discussion of ``{M}arkov chains for exploring posterior
  distributions".
\newblock \textit{The Annals of Statistics} \textbf{22} 1747--1757.

\bibitem[{Chatterjee and Lahiri(2011)}]{chat:lahi:2011}
\text{Chatterjee, A.} and \text{Lahiri, S.~N.} (2011).
\newblock Bootstrapping lasso estimators.
\newblock \textit{Journal of the American Statistical Association} \textbf{106}
  608--625.

\bibitem[{Diaconis et~al.(2014)Diaconis, Seiler and Holmes}]{diac:2014}
\text{Diaconis, P.}, \text{Seiler, C.} and \text{Holmes, S.} (2014).
\newblock Connections and extensions: A discussion of the paper by {G}irolami
  and {B}yrne.
\newblock \textit{Scandinavian Journal of Statistics} \textbf{41} 3--7.

\bibitem[{Fan and Li(2001)}]{fan:li:2001}
\text{Fan, J.} and \text{Li, R.} (2001).
\newblock {Variable selection via nonconcave penalized likelihood and its
  oracle properties}.
\newblock \textit{Journal of the American statistical Association} \textbf{96}
  1348--1360.

\bibitem[{Fan and Peng(2016)}]{fan:peng:2016}
\text{Fan, Y.} and \text{Peng, Q.} (2016).
\newblock {Inferring gene regulatory networks based on spline regression and
  Bayesian group lasso}.
\newblock In \textit{2016 17th IEEE/ACIS International Conference on Software
  Engineering, Artificial Intelligence, Networking and Parallel/Distributed
  Computing (SNPD)}.

\bibitem[{Flegal et~al.(2017)Flegal, Hughes, Vats and Dai}]{FlegalEtal:2017}
\text{Flegal, J.~M.}, \text{Hughes, J.}, \text{Vats, D.} and \text{Dai, N.}
  (2017).
\newblock mcmcse: Monte carlo standard errors for mcmc.
\newblock \textit{Riverside, CA and Minneapolis, MN. R package version}
  1.3--2.

\bibitem[{George and McCulloch(1993)}]{george:mcculloch:1993}
\text{George, E.~I.} and \text{McCulloch, R.~E.} (1993).
\newblock {Variable selection via Gibbs sampling}.
\newblock \textit{Journal of the American Statistical Association} \textbf{88}
  881--889.

\bibitem[{Ghosh and Tan(2015)}]{ghos:tan:2015}
\text{Ghosh, J.} and \text{Tan, A.} (2015).
\newblock {Sandwich algorithms for Bayesian variable selection}.
\newblock \textit{Computational Statistics \& Data Analysis} \textbf{81}
  76--88.

\bibitem[{Greenlaw et~al.(2017)Greenlaw, Szefer, Graham, Lesperance, Nathoo and
  Initiative}]{greenlawEtal:2017}
\text{Greenlaw, K.}, \text{Szefer, E.}, \text{Graham, J.}, \text{Lesperance,
  M.}, \text{Nathoo, F.~S.} and \text{Initiative, A. D.~N.} (2017).
\newblock {A Bayesian group sparse multi-task regression model for imaging
  genetics}.
\newblock \textit{Bioinformatics} \textbf{33} 2513--2522.

\bibitem[{Griffin and Brown(2010)}]{grifEtal:2010}
\text{Griffin, J.~E.} and \text{Brown, P.~J.} (2010).
\newblock Inference with normal-gamma prior distributions in regression
  problems.
\newblock \textit{Bayesian Analysis} \textbf{5} 171--188.

\bibitem[{Hefley et~al.(2017)Hefley, Hooten, Hanks, Russell and
  Walsh}]{hefleyEtal:2017}
\text{Hefley, T.~J.}, \text{Hooten, M.~B.}, \text{Hanks, E.~M.}, \text{Russell,
  R.~E.} and \text{Walsh, D.~P.} (2017).
\newblock {The Bayesian group lasso for confounded spatial data}.
\newblock \textit{Journal of Agricultural, Biological and Environmental
  Statistics} \textbf{22} 42--59.

\bibitem[{Hobert and Marchev(2008)}]{hobert:Marchev:2008}
\text{Hobert, J.~P.} and \text{Marchev, D.} (2008).
\newblock {A theoretical comparison of the data augmentation, marginal
  augmentation and PX-DA algorithms}.
\newblock \textit{The Annals of Statistics}  532--554.

\bibitem[{Hobert et~al.(2011)Hobert, Roy and Robert}]{hobe:roy:robe:2011}
\text{Hobert, J.~P.}, \text{Roy, V.} and \text{Robert, C.~P.} (2011).
\newblock {Improving the convergence properties of the data augmentation
  algorithm with an application to Bayesian mixture modeling}.
\newblock \textit{Statistical Science} \textbf{26} 332--351.

\bibitem[{Hooten and Hobbs(2015)}]{hooten:bobbs:2015}
\text{Hooten, M.~B.} and \text{Hobbs, N.} (2015).
\newblock {A guide to Bayesian model selection for ecologists}.
\newblock \textit{Ecological Monographs} \textbf{85} 3--28.

\bibitem[{Huang et~al.(2012)Huang, Breheny and Ma}]{huan:breh:ma:2012}
\text{Huang, J.}, \text{Breheny, P.} and \text{Ma, S.} (2012).
\newblock A selective review of group selection in high-dimensional models.
\newblock \textit{Statistical Science} \textbf{27}.

\bibitem[{Johnstone and Silverman(2004)}]{john:silv:2004}
\text{Johnstone, I.~M.} and \text{Silverman, B.~W.} (2004).
\newblock {Needles and straw in haystacks: Empirical Bayes estimates of
  possibly sparse sequences}.
\newblock \textit{The Annals of Statistics} \textbf{32} 1594--1649.

\bibitem[{Khare and Hobert(2011)}]{khar:hobe:2011}
\text{Khare, K.} and \text{Hobert, J.~P.} (2011).
\newblock A spectral analytic comparison of trace-class data augmentation
  algorithms and their sandwich variants.
\newblock \textit{The Annals of Statistics} \textbf{39} 2585--2606.

\bibitem[{Kyung et~al.(2010)Kyung, Gill, Ghosh and Casella}]{kyungEtal:2010}
\text{Kyung, M.}, \text{Gill, J.}, \text{Ghosh, M.} and \text{Casella, G.}
  (2010).
\newblock Penalized regression, standard errors, and {B}ayesian lassos.
\newblock \textit{Bayesian Analysis} \textbf{5} 369--411.

\bibitem[{Lee and Chen(2015)}]{lee:chen:2015}
\text{Lee, K.-J.} and \text{Chen, R.-B.} (2015).
\newblock {BSGS: Bayesian Sparse Group Selection}.
\newblock \textit{R Journal} \textbf{7}.

\bibitem[{Li et~al.(2015)Li, Wang, Li and Wu}]{liEtal:2015}
\text{Li, J.}, \text{Wang, Z.}, \text{Li, R.} and \text{Wu, R.} (2015).
\newblock {Bayesian group LASSO for nonparametric varying-coefficient models
  with application to functional genome-wide association studies}.
\newblock \textit{The Annals of Applied Statistics} \textbf{9} 640.

\bibitem[{Li and Lin(2010)}]{li:lin:2010}
\text{Li, Q.} and \text{Lin, N.} (2010).
\newblock {The Bayesian elastic net}.
\newblock \textit{Bayesian analysis} \textbf{5} 151--170.

\bibitem[{Liu et~al.(1994)Liu, Wong and Kong}]{junEtal:1994}
\text{Liu, J.~S.}, \text{Wong, W.~H.} and \text{Kong, A.} (1994).
\newblock {Covariance structure of the Gibbs sampler with applications to the
  comparisons of estimators and augmentation schemes}.
\newblock \textit{Biometrika} \textbf{81} 27--40.

\bibitem[{Liu and Wu(1999)}]{liu:wu:1999}
\text{Liu, J.~S.} and \text{Wu, Y.~N.} (1999).
\newblock Parameter expansion for data augmentation.
\newblock \textit{Journal of the American Statistical Association} \textbf{94}
  1264--1274.

\bibitem[{Livingstone et~al.(2016)Livingstone, Betancourt, Byrne and
  Girolami}]{livi:beta:byrn:giro:2016}
\text{Livingstone, S.}, \text{Betancourt, M.}, \text{Byrne, S.} and
  \text{Girolami, M.} (2016).
\newblock {On the geometric ergodicity of Hamiltonian Monte Carlo}.
\newblock \textit{arXiv preprint arXiv:1601.08057} .

\bibitem[{Livingstone and Zanella(2019)}]{sam:giac:2019}
\text{Livingstone, S.} and \text{Zanella, G.} (2019).
\newblock {On the robustness of gradient-based MCMC algorithms}.
\newblock \textit{arXiv preprint arXiv:1908.11812} .

\bibitem[{Lockhart et~al.(2014)Lockhart, Taylor, Tibshirani and
  Tibshirani}]{lock:2014}
\text{Lockhart, R.}, \text{Taylor, J.}, \text{Tibshirani, R.~J.} and
  \text{Tibshirani, R.} (2014).
\newblock A significance test for the lasso.
\newblock \textit{The Annals of Statistics} \textbf{42} 413.

\bibitem[{Marchev and Hobert(2004)}]{marc:hobe:2004}
\text{Marchev, D.} and \text{Hobert, J.~P.} (2004).
\newblock {Geometric ergodicity of van Dyk and Meng's algorithm for the
  multivariate Student's t model}.
\newblock \textit{Journal of the American Statistical Association} \textbf{99}
  228--238.

\bibitem[{Meng and Van~Dyk(1999)}]{meng:vand:1999}
\text{Meng, X.-L.} and \text{Van~Dyk, D.~A.} (1999).
\newblock Seeking efficient data augmentation schemes via conditional and
  marginal augmentation.
\newblock \textit{Biometrika} \textbf{86} 301--320.

\bibitem[{Mira(2001)}]{mira:2001}
\text{Mira, A.} (2001).
\newblock {Ordering and improving the performance of Monte Carlo Markov
  chains}.
\newblock \textit{Statistical Science}  340--350.

\bibitem[{Mitchell and Beauchamp(1988)}]{mitc:beau:88}
\text{Mitchell, T.~J.} and \text{Beauchamp, J.~J.} (1988).
\newblock Bayesian variable selection in linear regression.
\newblock \textit{Journal of the American Statistical Association} \textbf{83}
  1023--1032.

\bibitem[{Neal(2011)}]{neal:2011}
\text{Neal, R.~M.} (2011).
\newblock {MCMC using Hamiltonian dynamics}.
\newblock In \textit{Handbook of {M}arkov chain {Monte} {C}arlo} (S.~Brooks,
  A.~Gelman, G.~Jones and X.-L. Meng, eds.), chap.~5. CRC press, 113--162.

\bibitem[{Neville et~al.(2014)Neville, Ormerod and Wand}]{nevi:orme:wand:2014}
\text{Neville, S.~E.}, \text{Ormerod, J.~T.} and \text{Wand, M.} (2014).
\newblock Mean field variational bayes for continuous sparse signal shrinkage:
  pitfalls and remedies.
\newblock \textit{Electronic Journal of Statistics} \textbf{8} 1113--1151.

\bibitem[{Pal et~al.(2017)Pal, Khare and Hobert}]{palEtal:2017}
\text{Pal, S.}, \text{Khare, K.} and \text{Hobert, J.~P.} (2017).
\newblock {Trace class Markov chains for Bayesian inference with generalized
  double Pareto shrinkage priors}.
\newblock \textit{Scandinavian Journal of Statistics} \textbf{44} 307--323.

\bibitem[{Park and Casella(2008)}]{park:case:2008}
\text{Park, T.} and \text{Casella, G.} (2008).
\newblock The {B}ayesian lasso.
\newblock \textit{Journal of the American Statistical Association} \textbf{103}
  681--686.

\bibitem[{Plummer et~al.(2006)Plummer, Best, Cowles and
  Vines}]{plummerEtal:2006}
\text{Plummer, M.}, \text{Best, N.}, \text{Cowles, K.} and \text{Vines, K.}
  (2006).
\newblock {CODA: convergence diagnosis and output analysis for MCMC}.
\newblock \textit{R news} \textbf{6} 7--11.

\bibitem[{Qin and Hobert(2018)}]{qin:hobe:2018}
\text{Qin, Q.} and \text{Hobert, J.~P.} (2018).
\newblock {Wasserstein-based methods for convergence complexity analysis of
  MCMC with application to Albert and Chib's algorithm}.
\newblock \textit{arXiv preprint arXiv:1810.08826} .

\bibitem[{Qin and Hobert(2019)}]{qin:hobe:2019}
\text{Qin, Q.} and \text{Hobert, J.~P.} (2019).
\newblock {Convergence complexity analysis of Albert and Chib's algorithm for
  Bayesian probit regression}.
\newblock \textit{The Annals of Statistics} .

\bibitem[{Rajaratnam and Sparks(2015)}]{rajar:spar:2015}
\text{Rajaratnam, B.} and \text{Sparks, D.} (2015).
\newblock {MCMC}-based inference in the era of big data: A fundamental analysis
  of the convergence complexity of high-dimensional chains.
\newblock \textit{arXiv preprint arXiv:1508.00947} .

\bibitem[{Rajaratnam et~al.(2018)Rajaratnam, Sparks, Khare and
  Zhang}]{balaEtal:2018}
\text{Rajaratnam, B.}, \text{Sparks, D.}, \text{Khare, K.} and \text{Zhang, L.}
  (2018).
\newblock Uncertainty quantification for modern high-dimensional regression via
  scalable bayesian methods.
\newblock \textit{Journal of Computational and Graphical Statistics} .

\bibitem[{Raman et~al.(2009)Raman, Fuchs, Wild, Dahl and Roth}]{ramanEtal:2009}
\text{Raman, S.}, \text{Fuchs, T.~J.}, \text{Wild, P.~J.}, \text{Dahl, E.} and
  \text{Roth, V.} (2009).
\newblock {The Bayesian group-lasso for analyzing contingency tables}.
\newblock In \textit{Proceedings of the 26th Annual International Conference on
  Machine Learning}.

\bibitem[{Roberts and Rosenthal(1997)}]{robe:rose:1997}
\text{Roberts, G.~O.} and \text{Rosenthal, J.~S.} (1997).
\newblock Geometric ergodicity and hybrid {M}arkov chains.
\newblock \textit{Electronic Communications in Probability} \textbf{2} 13--25.

\bibitem[{Rose and Spiegel(2010)}]{rose:spiegel:2010}
\text{Rose, A.~K.} and \text{Spiegel, M.~M.} (2010).
\newblock {Cross-country causes and consequences of the 2008 crisis:
  International linkages and American exposure}.
\newblock \textit{Pacific Economic Review} \textbf{15} 340--363.

\bibitem[{Rosenthal et~al.(2015)Rosenthal, Rosenthal et~al.}]{rose:2015}
\text{Rosenthal, J.}, \text{Rosenthal, P.} \text{et~al.} (2015).
\newblock {Spectral bounds for certain two-factor non-reversible MCMC
  algorithms}.
\newblock \textit{Electronic Communications in Probability} \textbf{20}.

\bibitem[{Rosenthal(1995)}]{rose:1995}
\text{Rosenthal, J.~S.} (1995).
\newblock {Minorization conditions and convergence rates for Markov chain Monte
  Carlo}.
\newblock \textit{Journal of the American Statistical Association} \textbf{90}
  558--566.

\bibitem[{Scheetz et~al.(2006)Scheetz, Kim, Swiderski, Philp, Braun, Knudtson,
  Dorrance, DiBona, Huang, Casavant, Sheffield and Stone}]{scheetzEtal:2006}
\text{Scheetz, T.~E.}, \text{Kim, K.-Y.~A.}, \text{Swiderski, R.~E.},
  \text{Philp, A.~R.}, \text{Braun, T.~A.}, \text{Knudtson, K.~L.},
  \text{Dorrance, A.~M.}, \text{DiBona, G.~F.}, \text{Huang, J.},
  \text{Casavant, T.~L.}, \text{Sheffield, V.~C.} and \text{Stone, E.~M.}
  (2006).
\newblock {Regulation of gene expression in the mammalian eye and its relevance
  to eye disease}.
\newblock \textit{Proceedings of the National Academy of Sciences} \textbf{103}
  14429--14434.

\bibitem[{Simon et~al.(2013)Simon, Friedman, Hastie and
  Tibshirani}]{noahEtal:2013}
\text{Simon, N.}, \text{Friedman, J.}, \text{Hastie, T.} and \text{Tibshirani,
  R.} (2013).
\newblock A sparse group lasso.
\newblock \textit{Journal of Computational and Graphical Statistics}
  \textbf{22} 231--245.

\bibitem[{Tan and Hobert(2009)}]{tan:hobe:2009}
\text{Tan, A.} and \text{Hobert, J.~P.} (2009).
\newblock Block {G}ibbs sampling for {B}ayesian random effects models with
  improper priors: {C}onvergence and regeneration.
\newblock \textit{Journal of Computational and Graphical Statistics}
  \textbf{18} 861--878.

\bibitem[{Tibshirani et~al.(2005)Tibshirani, Saunders, Rosset, Zhu and
  Knight}]{robertEtal:2015}
\text{Tibshirani, R.}, \text{Saunders, M.}, \text{Rosset, S.}, \text{Zhu, J.}
  and \text{Knight, K.} (2005).
\newblock Sparsity and smoothness via the fused lasso.
\newblock \textit{Journal of the Royal Statistical Society, {\rm Series~B}}
  \textbf{67} 91--108.

\bibitem[{Tibshirani and Wang(2007)}]{tibshirani:pei:2007}
\text{Tibshirani, R.} and \text{Wang, P.} (2007).
\newblock {Spatial smoothing and hot spot detection for CGH data using the
  fused lasso}.
\newblock \textit{Biostatistics} \textbf{9} 18--29.

\bibitem[{Tibshirani et~al.(2016)Tibshirani, Taylor, Lockhart and
  Tibshirani}]{tibs:2016}
\text{Tibshirani, R.~J.}, \text{Taylor, J.}, \text{Lockhart, R.} and
  \text{Tibshirani, R.} (2016).
\newblock Exact post-selection inference for sequential regression procedures.
\newblock \textit{Journal of the American Statistical Association} \textbf{111}
  600--620.

\bibitem[{Van~Dyk and Meng(2001)}]{vand:meng:2001}
\text{Van~Dyk, D.~A.} and \text{Meng, X.-L.} (2001).
\newblock {The art of data augmentation}.
\newblock \textit{Journal of Computational and Graphical Statistics}
  \textbf{10} 1--50.

\bibitem[{Vats(2017)}]{vats:2017}
\text{Vats, D.} (2017).
\newblock {Geometric ergodicity of Gibbs samplers in Bayesian penalized
  regression models}.
\newblock \textit{Electronic Journal of Statistics} \textbf{11} 4033--4064.

\bibitem[{Vats et~al.(2019)Vats, Flegal and Jones}]{vats:fleg:jone:2019}
\text{Vats, D.}, \text{Flegal, J.~M.} and \text{Jones, G.~L.} (2019).
\newblock Multivariate output analysis for markov chain monte carlo.
\newblock \textit{Biometrika} \textbf{106} 321--337.

\bibitem[{Wilks(1932)}]{wilk:1932}
\text{Wilks, S.~S.} (1932).
\newblock Certain generalizations in the analysis of variance.
\newblock \textit{Biometrika}  471--494.

\bibitem[{Xu and Ghosh(2015)}]{xf:ghosh:2015}
\text{Xu, X.} and \text{Ghosh, M.} (2015).
\newblock Bayesian variable selection and estimation for group lasso.
\newblock \textit{Bayesian Analysis} \textbf{10} 909--936.

\bibitem[{Yang and Rosenthal(2017)}]{yang:rose:2017}
\text{Yang, J.} and \text{Rosenthal, J.~S.} (2017).
\newblock Complexity results for {MCMC} derived from quantitative bounds.
\newblock \textit{arXiv preprint arXiv:1708.00829} .

\bibitem[{Yang and Zou(2015)}]{yang:zou:2015}
\text{Yang, Y.} and \text{Zou, H.} (2015).
\newblock A fast unified algorithm for solving group-lasso penalize learning
  problems.
\newblock \textit{Statistics and Computing} \textbf{25} 1129--1141.

\bibitem[{Yuan and Lin(2006)}]{ming:yi:2006}
\text{Yuan, M.} and \text{Lin, Y.} (2006).
\newblock Model selection and estimation in regression with grouped variables.
\newblock \textit{Journal of the Royal Statistical Society, {\rm Series~B}}
  \textbf{68} 49--67.

\bibitem[{Zhang et~al.(2014)Zhang, Baladandayuthapani, Mallick, Manyam,
  Thompson, Bondy and Do}]{zhangEtal:2014}
\text{Zhang, L.}, \text{Baladandayuthapani, V.}, \text{Mallick, B.~K.},
  \text{Manyam, G.~C.}, \text{Thompson, P.~A.}, \text{Bondy, M.~L.} and
  \text{Do, K.-A.} (2014).
\newblock {Bayesian hierarchical structured variable selection methods with
  application to molecular inversion probe studies in breast cancer}.
\newblock \textit{Journal of the Royal Statistical Society: Series C (Applied
  Statistics)} \textbf{63} 595--620.

\end{thebibliography}
\end{document}